\documentclass[11pt]{article} 
\usepackage{amsmath}
\usepackage{amsfonts,amssymb,latexsym}

\begin{document}

\newcommand{\nl}{\nonumber\\}
\newcommand{\nnl}{\nl[6mm]}
\newcommand{\nle}{\nl[-2.5mm]\\[-2.5mm]}
\newcommand{\nlb}[1]{\nl[-2.0mm]\label{#1}\\[-2.0mm]}
\newcommand{\ab}{\allowbreak}

\renewcommand{\leq}{\leqslant}              
\renewcommand{\geq}{\geqslant}

\renewcommand{\theequation}{\thesection.\arabic{equation}}
\let\ssection=\section
\renewcommand{\section}{\setcounter{equation}{0}\ssection}

\newcommand{\be}{\bes}
\newcommand{\ee}{\ees}
\newcommand{\bes}{\begin{eqnarray}}
\newcommand{\ees}{\end{eqnarray}}
\newcommand{\eens}{\nonumber\end{eqnarray}}
\newcommand{\barr}{\begin{array}}
\newcommand{\earr}{\end{array}}

\renewcommand{\/}{\over}
\renewcommand{\d}{\partial}
\newcommand{\Dslash}{\hbox{$D\kern-2.4mm/\,$}}
\newcommand{\dd}[1]{\ab\delta/\delta {#1}}
\newcommand{\ddt}{{d\/dt}}

\newcommand{\no}[1]{{\,:\kern-0.7mm #1\kern-1.2mm:\,}}

\newcommand{\mm}{{\mathbf m}}
\newcommand{\nn}{{\mathbf n}}
\newcommand{\cmm}{{,\mm}}
\newcommand{\cnn}{{,\nn}}

\newcommand{\xmu}{\xi^\mu}
\newcommand{\ynu}{\eta^\nu}
\newcommand{\qsmu}{q^*_\mu}
\newcommand{\qsnu}{q^*_\nu}
\newcommand{\psmu}{p_*^\mu}

\newcommand{\half}{{1\/2}}
\newcommand{\quart}{{1\/4}}
\newcommand{\tpi}{{1\/2\pi i}}
\newcommand{\bra}[1]{\big{\langle}#1\big{|}}
\newcommand{\ket}[1]{\big{|}#1\big{\rangle}}

\newcommand{\da}{\d_\alpha}
\newcommand{\db}{\d_\beta}
\newcommand{\dc}{\d_\gamma}
\newcommand{\za}{\zeta_a}
\newcommand{\zb}{\zeta_b}
\newcommand{\zc}{\zeta_c}
\newcommand{\Mab}{M^{\alpha\beta}}
\newcommand{\Kab}{K_{\alpha\beta}}

\newcommand{\fa}{\phi^\alpha}
\newcommand{\fb}{\phi^\beta}
\newcommand{\pa}{\pi_\alpha}
\newcommand{\pb}{\pi_\beta}
\newcommand{\Ea}{\EE_\alpha}
\newcommand{\Eb}{\EE_\beta}

\newcommand{\fs}{\phi^*}
\newcommand{\ps}{\pi_*}
\newcommand{\fsa}{\phi^*_\alpha}
\newcommand{\fsb}{\phi^*_\beta}
\newcommand{\fsc}{\phi^*_\gamma}
\newcommand{\psa}{\pi_*^\alpha}
\newcommand{\psb}{\pi_*^\beta}
\newcommand{\fsi}{\phi^*_i}

\newcommand{\fA}{\phi^A}
\newcommand{\fB}{\phi^B}
\newcommand{\pA}{\pi_A}
\newcommand{\wfA}{\bar\phi^A}
\newcommand{\wfB}{\bar\phi^B}

\newcommand{\As}{A_*}
\newcommand{\Es}{E^*}
\newcommand{\wA}{{\bar A}}
\newcommand{\wE}{{\bar E}}
\newcommand{\wAs}{{\bar A}{}_*}
\newcommand{\wEs}{{\bar E}{}^*}
\newcommand{\wc}{{\bar c}}
\newcommand{\wzeta}{{\bar\zeta}}
\newcommand{\wF}{{\bar F}}

\newcommand{\wf}{{\bar \phi}}
\renewcommand{\wp}{{\bar \pi}}
\newcommand{\wfs}{{\bar \phi}{}^*}
\newcommand{\wps}{{\bar \pi}_*}

\newcommand{\fm}{\phi_\cmm}
\newcommand{\fn}{\phi_\cnn}
\newcommand{\pim}{\pi^\cmm}
\newcommand{\pin}{\pi^\cnn}
\newcommand{\fsm}{\phi^*_\cmm}
\newcommand{\fsn}{\phi^*_\cnn}
\newcommand{\psm}{\pi_*^\cmm}
\newcommand{\psn}{\pi_*^\cnn}

\renewcommand{\fam}{\phi^\alpha_\cmm}
\newcommand{\fbn}{\phi^\beta_\cnn}
\newcommand{\pam}{\pi_\alpha^\cmm}
\newcommand{\pbn}{\pi_\beta^\cnn}
\newcommand{\zam}{\zeta_{a\cmm}}
\newcommand{\zcm}{\zeta_{c\cmm}}
\newcommand{\cam}{c^a_\cmm}
\newcommand{\wfam}{{\bar \phi}{}^\alpha_\cmm}
\newcommand{\wfbn}{{\bar \phi}{}^\beta_\cnn}
\newcommand{\wpam}{{\bar \pi}{}_\alpha^\cmm}

\newcommand{\fAm}{\fA_\cmm}
\newcommand{\fAn}{\fA_\cnn}
\newcommand{\fBn}{\fB_\cnn}
\newcommand{\pAm}{\pA^\cmm}
\newcommand{\pAn}{\pA^\cnn}
\newcommand{\wfAm}{\bar\fA_\cmm}

\newcommand{\wfm}{{\bar \phi}_\cmm}
\newcommand{\wfn}{{\bar \phi}_\cnn}
\newcommand{\wpm}{{\bar \pi}^\cmm}
\newcommand{\wfsm}{\bar \phi^*_\cmm}
\newcommand{\wfsn}{\bar \phi^*_\cnn}
\newcommand{\wpsm}{\bar \pi_*^\cmm}
\newcommand{\wpsn}{\bar \pi_*^\cnn}

\newcommand{\wfa}{{\bar \phi}{}^\alpha}
\newcommand{\wfb}{{\bar \phi}{}^\beta}
\newcommand{\wpa}{\bar \pi_\alpha}
\newcommand{\wpb}{\bar \pi_\beta}

\newcommand{\fsam}{\phi^*_{\alpha,\mm}}
\newcommand{\fsbn}{\phi^*_{\beta,\nn}}
\newcommand{\psam}{\pi_*^{\alpha,\mm}}
\newcommand{\psbn}{\pi_*^{\beta,\nn}}
\newcommand{\Eam}{\EE_{\alpha,\mm}}

\newcommand{\wfsa}{\bar \phi^*_\alpha}
\newcommand{\wpsa}{\bar \pi_*^\alpha}
\newcommand{\wpsb}{\bar \pi_*^\beta}
\newcommand{\wfsam}{\bar \phi^*_{\alpha,\mm}}
\newcommand{\wfsan}{\bar \phi^*_{\alpha,\nn}}
\newcommand{\wfsbn}{\bar \phi^*_{\beta,\nn}}
\newcommand{\wpsam}{\bar \pi_*^{\alpha,\mm}}
\newcommand{\wpsan}{\bar \pi_*^{\alpha,\nn}}
\newcommand{\wpsbm}{\bar \pi_*^{\beta,\mm}}
\newcommand{\wpsbn}{\bar \pi_*^{\beta,\nn}}

\newcommand{\dam}{\d_\alpha^\mm}
\newcommand{\dbn}{\d_\beta^\nn}

\newcommand{\ord}{o}
\newcommand{\ordg}{\varsigma}

\newcommand{\Np}[1]{{N+p\choose N #1}}
\newcommand{\Npr}{{N+p-r\choose N-r}}
\newcommand{\ritwo}{(-)^i {r-2\choose i-2}}
\newcommand{\rione}{(-)^i {r-1\choose i-1}}

\newcommand{\si}{\sigma}
\newcommand{\eps}{\epsilon}
\newcommand{\dlt}{\delta}
\newcommand{\om}{\omega}
\newcommand{\al}{\alpha}
\newcommand{\bt}{\beta}
\newcommand{\gm}{\gamma}
\newcommand{\ka}{\kappa}
\newcommand{\la}{\lambda}
\newcommand{\vth}{\vartheta}
\renewcommand{\th}{\theta}
\newcommand{\rep}{\varrho}

\newcommand{\vect}{{\mathfrak{vect}}}
\newcommand{\map}{{\mathfrak{map}}}
\newcommand{\dmap}{\vect(N)\ltimes \map(N,\oj)}

\newcommand{\im}{{\rm im}\ }
\newcommand{\ext}{{\rm ext}\ }
\newcommand{\e}{{\rm e}}
\renewcommand{\div}{{\rm div}}
\newcommand{\afn}{{\rm afn\,}}
\newcommand{\til}{{\tilde{\ }}}

\newcommand{\eikx}{\e^{ik\cdot x}}
\newcommand{\dNx}{{d^N \kern-0.4mm x}}
\newcommand{\dNk}{{d^N \kern-0.4mm k}}
\newcommand{\dNxp}{{d^N \kern-0.4mm x'}}
\newcommand{\dNxb}{{d^N \kern-0.4mm x''}}
\newcommand{\dNkp}{{d^N \kern-0.4mm k'}}
\newcommand{\dFx}{{d^4 \kern-0.4mm x}}

\newcommand{\summ}[1]{\sum_{|\mm|\leq #1}}
\newcommand{\sumn}[1]{\sum_{|\nn|\leq #1}}
\newcommand{\summnp}{\sum_{|\mm|\leq|\nn|\leq p }}
\newcommand{\sumnmp}[1]{\sum_{|\nn|\leq|\mm|\leq p#1 }}
\newcommand{\summp}{\summ{p}}

\newcommand{\dmu}{{\d_\mu}}
\newcommand{\dnu}{{\d_\nu}}

\newcommand{\larroww}[1]{{\ \stackrel{#1}{\longleftarrow}\ }}
\newcommand{\rarroww}[1]{{\ \stackrel{#1}{\longrightarrow}\ }}
\newcommand{\intdm}{\sum_{m=-\infty}^\infty }

\newcommand{\repi}{\rep^{(i)}}
\newcommand{\Mi}{M^{(i)}}
\newcommand{\mri}{(-)^i {r\choose i}}

\newcommand{\mndmn}{{\mm\choose\nn}\d_{\mm-\nn}}

\newcommand{\trrep}{{\rm tr}_{\rep}\kern0.7mm}
\newcommand{\trM}{{\rm tr}_{M}\kern0.7mm}
\newcommand{\tr}{{\rm tr}}
\newcommand{\oj}{{\mathfrak g}}
\newcommand{\g}{{\mathfrak g}}
\newcommand{\hh}{{\mathfrak h}}
\newcommand{\su}{{\mathfrak su}}
\newcommand{\uu}{{\mathfrak u}}

\newcommand{\ssu}{\su(3)\!\oplus\! \su(2)\!\oplus\! \uu(1)}

\renewcommand{\L}{{\cal L}}
\newcommand{\J}{{\cal J}}
\newcommand{\D}{{\cal D}}
\newcommand{\U}{{\cal U}}
\newcommand{\N}{{\cal N}}
\newcommand{\OO}{{\cal O}}
\newcommand{\QQ}{{\cal Q}}
\newcommand{\PP}{{\cal P}}
\newcommand{\EE}{{\cal E}}
\newcommand{\FF}{{\cal F}}
\newcommand{\HH}{{\cal H}}
\newcommand{\II}{{\cal I}}
\newcommand{\GG}{{\cal G}}

\newcommand{\cl}{{cl}}
\newcommand{\qm}{{qm}}
                                            
\newcommand{\TT}{{\mathbb T}}
\newcommand{\RR}{{\mathbb R}}
\newcommand{\CC}{{\mathbb C}}
\newcommand{\ZZ}{{\mathbb Z}}
\newcommand{\NN}{{\mathbb N}}

\title{{Manifestly covariant canonical quantization III: Gravity,
locality, and diffeomorphism anomalies in four dimensions}}

\author{T. A. Larsson \\
Vanadisv\"agen 29, S-113 23 Stockholm, Sweden\\
email: thomas.larsson@hdd.se}

\maketitle 

\begin{abstract} 
The recently introduced manifestly covariant canonical quantization 
scheme is applied to gravity. New diffeomorphism anomalies 
generating a multi-dimensional generalization of the Virasoro algebra
arise. This does not contradict theorems about the non-existence of
gravitational anomalies in four dimensions, because the relevant 
cocycles depend on the observer's spacetime trajectory, which is
ignored in conventional field theory. Rather than being inconsistent,
these anomalies are necessary to obtain a {\em local} theory of
quantum gravity.
\end{abstract}

\vskip 3 cm
PACS (2003): 02.20.Tw, 03.70.+k, 04.60.Ds

\bigskip
Keywords: Covariant canonical quantization, History phase space,
Multi-dimensional Virasoro algebra, Diffeomorphism anomalies.

\newpage

\section{Introduction}
\label{sec:Intro}

The construction of a quantum theory of gravity is the outstanding
problem in physics today \cite{Car01,Smo03}. To make some progress, it
is useful to consider the key concepts underlying its constituent
subtheories.

Quantum field theory (QFT) rests on three conceptual pillars:
\begin{enumerate}
\item
Quantum mechanics.
\item
Lorentz invariance, i.e. special relativity.
\item
Cluster decomposition.
\end{enumerate}
General relativity rests on two pillars:
\begin{enumerate}
\item
General covariance, i.e. diffeomorphism invariance (without
background fields).
\item
Locality.
\end{enumerate}
Since Lorentz invariance may be viewed as a special case of general
covariance, and cluster decomposition is a statement about locality,
this suggests that quantum gravity should be based on three principles:
\begin{enumerate}
\item
Quantum mechanics.
\item
Diffeomorphism invariance.
\item
Locality.
\end{enumerate}
Alas, it is well known that diffeomorphism invariance and locality
are incompatible in quantum theory \cite{dW67}; ``there are no
local observables in quantum gravity''. Hence the quest for a
theory satisfying all three principles above appears futile.

However, there is a loophole. Proper diffeomorphism invariance is
a gauge symmetry, i.e. a mere redundancy of the description, but
the situation might be changed in the presence of anomalies. In
the treatment of quantum black holes in three dimensions
\cite{Car05}, diffeomorphism anomalies (or ``would-be
diffeomorphisms'') on the boundary are responsible for entropy.
More to the point, it is well known in the context of conformal
field theory (CFT) \cite{FMS96}, that it is possible to combine
infinite spacetime symmetries with locality, in the sense of
correlation functions depending on separation, but only in the
presence of an anomaly. Since the infinite conformal group in two
dimensions is isomorphic to (twice) the diffeomorphism group in
one dimension, all results in CFT apply to one-dimensional
diffeomorphism-invariant QFT as well. This is a kinematical
statement on the level of fields, correlation functions and Ward
identities, which does not involve dynamics and hence it is
independent of whether diffeomorphisms are gauge transformations
or not.

Although it has been suggested before that anomalies could be
responsible for gauge symmetry breaking \cite{JR85}, it would seem that
this could not be relevant to quantum gravity, since no pure
diffeomorphism anomalies exist in four dimensions \cite{Bon86}; see also
\cite{Wein96}, chapter 22. Or do they? Anomalies manifest themselves as
extensions of the constraint algebra, and the diffeomorphism algebra in
any number of dimensions certainly admits extensions; in fact, all
extensions by modules of tensor fields were classified by Dzhumadildaev
\cite{Dzhu96}. In particular, there is a higher-dimensional analogue of
the Virasoro algebra, whose representation theory was developed in
\cite{Lar98,Lar01,Lar02,RM94}; see \cite{Lar03} for a recent review.

The key idea is the introduction of the
observer's trajectory in spacetime. In non-relativistic quantum
mechanics, observation is a complicated, non-local process which assigns
numbers to experiments. However, in a relativistic theory, a process
must be localized; it happens {\em somewhere}. In order to maintain
locality, we must assign the process of observation to some definite
event in spacetime. As time proceeds, the observer (or detector or test
particle) traces out a curve in spacetime. Like the quantum
fields, the observer's trajectory should be treated as a material,
quantized object; it has a conjugate momentum, it is represented on the
Hilbert space, etc.

To consider the observer's trajectory as a material object is certainly
a very small modification, which does not introduce any new physical
ideas, and one may wonder why it should be important. The reason is that
it makes it possible for new types of anomalies to arise. If the field
theory has a gauge symmetry of Yang-Mills type, there is a gauge anomaly
proportional to the quadratic Casimir, and the gauge (or current)
algebra becomes a higher-dimensional generalization of affine Kac-Moody
algebras. Similarly, a general-covariant theory, in any number of
dimensions, acquires a diffeomorphism anomaly, which is described by the
above-mentioned multi-dimensional Virasoro algebra. The reason
why these anomalies can not be seen in conventional QFT, without
explicit reference to the observer, is that the relevant cocycles are
functionals of the observer's trajectory. If this trajectory has not
been introduced, it is of course impossible to write down the relevant
anomalies.

The canonical formalism is not very well suited for quantization of
relativistic theories, because the foliation of spacetime into fixed
time slices breaks manifest covariance. In fact, canonical quantization
proceeds in two steps, both of which violate covariance:
\begin{enumerate}
\item
Replace Poisson brackets by commutators. Phase space is defined 
non-covariantly on a space-like surface.
\item
Represent the resulting Heisenberg algebra on a Hilbert space with
energy bounded from below. The Hamiltonian refers to a
privileged time direction.
\end{enumerate}
To remedy these problems, a novel quantization scheme has recently been
proposed, called manifest covariant canonical quantization (MCCQ)
\cite{Lar04,Lar05a}. In the present paper, which is the last in this
series, we finally apply MCCQ to general relativity.
The two sources of non-covariance are addressed as follows:
\begin{enumerate}
\item
Phase space points are identified with histories which solve the 
dynamics. We quantize in the phase space of arbitrary histories first,
and impose dynamics as a first-class constraint afterwards.
\item
The observer's trajectory is introduced as a base point for a Taylor
expansion of all fields. The Hamiltonian is identified with the
generator of rigid translations parallel to the observer's trajectory.
The dynamical variables, i.e. the Taylor coefficients, live on this
trajectory and are hence causally related. The notion of spacelike
separation only arises implicitly by taking the (problematic) limit 
of infinite Taylor series.
\end{enumerate}

History methods have recently been advocated by Savvidou and Isham
\cite{Ish95,Sav99,Sav04}; in particular, the last reference contains a
very good summary of the conceptual problems involved in non-covariant
canonical quantization. Their formalism differs in details from MCCQ,
e.g. because they do not use cohomological methods. There is also a
substantial difference, namely that the observer's trajectory is not
introduced, and hence they can not see any diffeomorphism anomalies.

\section{Anomalies, consistency, locality, and unitarity}
\label{sec:Locality}

At this point it is necessary to discuss the issue of gauge anomalies
and consistency, in particular unitarity. Gauge and diffeomorphism
anomalies are usually considered as a sign of inconsistency and should
therefore be cancelled \cite{NAG85}. There is ample evidence that this
is the correct prescription for conventional gauge anomalies, arising
from chiral fermions coupled to gauge fields. However, the anomalies
discussed in this paper are of a completely different type, depending
on the observer's trajectory, and intuition derived from conventional
anomalies needs not apply. In fact, one can give a very simple
algebraic argument why conventional anomalies must vanish
\cite{Lar05b}, and this argument does not apply to the
observer-dependent anomalies considered here.

A quantum theory is defined by a Hilbert space and a Hamiltonian which
generates time evolution. The main conditions for consistency are
unitarity and lack of infinities. If the theory has some symmetries,
these must be realized as unitary operators acting on the Hilbert space
as well. In particular, if time translation is included among the
symmetries, which is the case for the Poincar\'e and diffeomorphism
algebras, a unitary representation of the symmetry algebra is usually
enough for consistency. {F}rom this viewpoint, there is a 1-1
correspondence between general-covariant QFTs and unitary
representations of the diffeomorphism group on a conventional Hilbert
space. Namely, given the QFT, its Hilbert space carries a unitary
representation of the diffeomorphism group. Conversely, if we have a
unitary representation of the diffeomorphism group, the Hilbert space on
which it acts can be interpreted as the Hilbert space of some
general-covariant QFT.

Unfortunately, this observation is not so powerful, because no
non-trivial, unitary, lowest-energy irreps of the diffeomorphism algebra
are known except in one dimension. Nevertheless, we are able to make
some very general observations.
Assume that some algebra $\g$ has a unitary representation $R$
and a subalgebra $\hh$. Then the restriction of $R$ to $\hh$ is still
unitary, and this must hold for every subalgebra $\hh$ of $\g$. In
particular, let $\g = \vect(N)$ be the diffeomorphism algebra in $N$
dimensions and $\hh = \vect(1)$ the diffeomorphism algebra in one
dimension. There are infinitely many such subalgebras, and the restriction
of $R$ to each and every one of them must be unitary. Fortunately, the
unitary irreps of the diffeomorphism algebra in one dimension are known.
The result is that the only proper unitary irrep is the trivial one, but
there are many unitary irreps with a diffeomorphism anomaly. {F}rom this
it follows that the trivial representation is the only unitary
representation also in $N$ dimensions.``There are no local observables
in quantum gravity''.

However, there is one well-known case where we know how to combine
locality and infinite spacetime symmetry with quantum theory: conformal 
field theory (CFT). Locality means that 
the correlation functions depend on separation. For two points $z$ and $w$
in $\RR$ or $\CC$, the correlator is
\be
  \langle \phi(z) \phi(w) \rangle \sim {1\/(z-w)^{2h}} + more,
\label{correlator}
\ee
where $more$ stands for less singular terms when $z \to w$. That the
correlation function has this form is a diffeomorphism-invariant 
statement. The
$more$ terms will change under an arbitary diffeomorphism, but the
leading singularity will always have the same form, and in particular
the anomalous dimension $h$ is well defined.

We can phrase this slightly differently. The short-distance
singularity only depends on two points being infinitesimally close.
This is good, because we cannot determine the finite distance between
two points without knowing about the metric. General relativity does
not have a background metric structure, but it does have a background
differentiable structure (locally at least), and that is enough for
defining anomalous dimensions. Diffeomorphisms move points around, but
they do not separate two points which are infinitesimally close.

The relevant algebra in CFT is not really the one-dimensional
diffeomorphism algebra (or the two-dimensional conformal algebra), but 
rather its central extension known as the Virasoro algebra:
\be
  [L_m, L_n] = (n-m) L_{m+n} - {c\/12} (m^3 - m) \delta_{m+n}.
\ee
A lowest-energy representation is characterized by a vacuum satisfying
\be
  L_0 \ket{0} = h \ket{0}, 
  \qquad L_m \ket{0} = 0\ \hbox{for all $m < 0$.}
\ee
In particular, the lowest $L_0$ eigenvalue can be identified with the
anomalous dimension $h$ in the correlation function (\ref{correlator}).
This means that locality, in the sense of correlation functions
depending on separation, requires that $h > 0$. It is well known
\cite{FMS96} that unitarity either implies that
\be
c = 1 - {6\/m(m+1)}, \qquad
h = h_{rs}(m) = {[(m+1)r-ms]^2-1\/4m(m+1)},
\ee
where $m\geq2$ and $1\leq r<m, 1\leq s<r$ are positive integers,
or that $c\geq1$, $h\geq0$. In particular, the central extension $c$ is
non-zero for any non-trivial, unitary irrep with $h \neq 0$. This leads
to the important observation:

\bigskip
\begin{tabular}{|l|}
\hline
Locality and unitarity are compatible with diffeomorphism (and \\
local conformal) symmetry only in the presence of an anomaly.  \\
\hline
\end{tabular}
\bigskip

This is true in higher dimensions as well. Consider the correlator
$\langle\phi(x) \phi(y)\rangle$, where $x$ and $y$ are points in $\RR^N$.
We could take some one-dimensional curve $q(t)$ passing through $x$ and
$y$, such that $x = q(t)$ and $y = q(t')$. Then the short-distance
behaviour is of the form
\be
  \langle \phi(x) \phi(y) \rangle \sim {1\/(t - t')^{2h}} + more,
\ee
and $h$ is independent of the choice of curve, provided that it is 
sufficiently regular. The subalgebra of $\vect(N)$ which preserves 
$q(t)$ is a Virasoro algebra, so $h > 0$ implies that $c > 0$.

This observation is completely standard in the application of CFT to
statistical physics in two dimensions. The simplest example of a unitary
model is the Ising model, which consists of three irreps, with $c = 1/2$
and $h = 0$, $h = 1/16$, and $h = 1/2$. The Ising model is perfectly
consistent despite the anomaly, both mathematically (unitarity), and more
importantly physically (it is realized in nature, in soft condensed
matter systems). The standard counter-argument is that infinite conformal
symmetry in condensed matter is not a gauge symmetry, but rather an
anomalous global symmetry. However, if we could take the classical limit
of such a system, the conformal symmetry would seem to be a gauge
symmetry. Namely, the anomaly vanishes in the classical limit, and we can
write down a classical BRST operator which is nilpotent, and the symmetry
is gauge on the classical level. There is no classical way to distinguish
between such a ``fake'' gauge symmetry and a genuine gauge symmetry which
extends to the quantum level. 

More generally, let us assume that we have some phase space, and a Lie
algebra $\g$ with generators $J_a$, satisfying
\be
  [J_a, J_b] = f_{ab}{}^c J_c,
\label{oj}
\ee
acts on this phase space. The Einstein convention is used;
repeated indices, one up and one down, are implicitly summed over. 
If the bracket with the Hamiltonian gives us
a new element in $\g$,
\be
  [J_a, H] = C^b_a J_b,
\ee
we say that $\g$ is a {\em symmetry} of the Hamiltonian system. If $\g$ in
addition contains arbitary functions of time, the symmetry is a {\em gauge
symmetry}. In this case, a solution to Hamilton's equations depends on
arbitrary functions of time and is thus not fully specified by the
positions and momenta at time $t = 0$. The standard example is
electromagnetism, where the zeroth component $A_0$ of the vector
potential is arbitrary, because its canonical momentum $F^{00} = 0$. An
arbitary time evolution is of course not acceptable. The reason why
this seems to happen is that a gauge symmetry is a redundancy of the
description; the true dynamical degrees of freedom are fewer than what
one na\"\i vely expects. In electromagnetism, the gauge potential has 
four components but the photon has only two polarizations.

There are various ways to handle quantization of gauge systems. One is to
eliminate the gauge degrees of freedom first and then quantize. This is
cumbersome and it is usually preferable to quantize first and eliminate
the gauge symmetries afterwards. The simplest way
is to require that the gauge generators annihilate physical states,
\be
  J_a \ket{phys} = 0,
\ee
and also that two physical states are equivalent if the differ by some
gauge state, $J_a |>$. This procedure produces a Hilbert space of
physical states.

However, one thing may go wrong. Upon quantization, a symmetry may
acquire some quantum corrections, so that $\g$ is replaced by
\be
  [J_a, J_b] = f_{ab}{}^c J_c + \hbar D_{ab} + O(\hbar^2).
\ee
The operator $D_{ab}$ is called an {\em anomaly}. We can also have 
anomalies of the type
\be
  [J_a, H] = C_a^b J_b + \hbar E_a + O(\hbar^2).
\ee
If we now try to keep the definition of a physical state, we see that
we must also demand that
\be
  D_{ab} \ket{phys} = 0.
\ee
This implies further reduction of the Hilbert space. In the case that
$D_{ab}$ is invertible, there are no physical states at all,
so the Hilbert space is empty. However, this
does not necessarily mean that the anomaly by itself is inconsistent,
only that our definition of physical states is. In the presence of an
anomaly, additional states become physical. So our Hilbert space
becomes larger, containing some, or even all, of the previous gauge
degrees of freedom. A gauge anomaly implies that the gauge symmetry is 
broken on the quantum level.

It is important to realize that such a ``fake'' gauge symmetry may well
be consistent. The Virasoro algebra is obviously anomalous, with the
central charge playing the role of the $D_{ab}$, and still it has unitary
representations with non-zero $c$. Of course, a gauge anomaly {\em may}
be inconsistent, if the anomalous algebra does not possess any unitary
representations. This is apparently what happens for the chiral-fermion
type anomaly which is relevant e.g. in the standard model.

\section{Multi-dimensional Virasoro algebra}
\label{sec:Virasoro}

All non-trivial, unitary, lowest-energy irreps of the diffeomorphism
algebra are anomalous, in any number of dimensions. This is well-known in
one dimension, where the diffeomorphism algebra acquires an extension
known as the Virasoro algebra. It is also true in several dimensions,
which one proves by considering the restriction to the many Virasoro
subalgebras living on lines in spacetime. This is perhaps rather 
surprising, in view of the following two no-go theorems:
\begin{itemize}
\item
The diffeomorphism algebra has no central extension except in one 
dimension.
\item
In field theory, there are no pure gravitational anomalies in four 
dimension.
\end{itemize}
However, the assumptions in these no-go theorems are too strong; the
Virasoro extension is not central except in one dimension, and one needs
to go slightly beyond field theory by explicitly specifying where 
observation takes place.

To make contact with the Virasoro algebra in its most familiar form, we
describe its multi-dimensional sibling in a Fourier basis on the
$N$-dimensional torus. Recall first that the algebra of diffeomorphisms on
the circle, $\vect(1)$, has generators
\be
L_m = -i \exp(imx) {d\/d x},
\label{Lm}
\ee
where $x \in S^1$.
$\vect(1)$ has a central extension, the Virasoro algebra:
\be
[L_m, L_n] = (n-m)L_{m+n} - {c\/12} (m^3-m) \delta_{m+n},
\label{Vir}
\ee
where $c$ is a c-number known as the central charge or conformal
anomaly. This means that the Virasoro algebra is a Lie 
algebra; anti-symmetry and the Jacobi identities still hold. The
term linear in $m$ is unimportant, because it can be removed by a
redefinition of $L_0$. The cubic term $m^3$ is a non-trivial 
extension which cannot be removed by any redefinition.

The generators (\ref{Lm}) immediately generalize to vector fields on the
$N$-dimensional torus:
\be
L_\mu(m) = -i \exp(i m_\rho x^\rho) \dmu,
\ee
where $x = (x^\mu)$, $\mu = 1, 2, ..., N$ is a point in 
$N$-dimensional space and $m = (m_\mu) \in \ZZ^N$. 
These operators generate the algebra $\vect(N)$:
\be
[L_\mu(m), L_\nu(n)] = n_\mu L_\nu(m+n) - m_\nu L_\mu(m+n).
\ee
The question is now whether the Virasoro extension, i.e. the 
$m^3$ term in (\ref{Vir}), also generalizes to higher dimensions.

Rewrite the ordinary Virasoro algebra (\ref{Vir}) as
\bes
[L_m, L_n] &=& (n-m)L_{m+n} + c m^2 n S_{m+n}, \nl
{[}L_m, S_n] &=& (n+m)S_{m+n}, \nle
{[}S_m, S_n] &=& 0, \nl
m S_m &\equiv& 0.
\eens
It is easy to see that the two formulations of the Virasoro algebra are
equivalent (the linear cocycle has been absorbed into a redefinition of
$L_0$). The second formulation immediately generalizes to $N$
dimensions. The defining relations are
\bes
[L_\mu(m), L_\nu(n)] &=& n_\mu L_\nu(m+n) - m_\nu L_\mu(m+n) \nl 
&&  + (c_1 m_\nu n_\mu + c_2 m_\mu n_\nu) m_\rho S^\rho(m+n), \nl
{[}L_\mu(m), S^\nu(n)] &=& n_\mu S^\nu(m+n)
 + \delta^\nu_\mu m_\rho S^\rho(m+n), 
\nlb{mVir}
{[}S^\mu(m), S^\nu(n)] &=& 0, \nl
m_\mu S^\mu(m) &\equiv& 0.
\eens
This is an extension of $\vect(N)$ by the abelian ideal with basis 
$S^\mu(m)$. 
Geometrically, we can think of $L_\mu(m)$ as a vector field 
and $S^\mu(m) = \eps^{\mu\nu_2..\nu_N} \ab S_{\nu_2..\nu_N}(m)$ 
as a dual one-form (and $S_{\nu_2..\nu_N}(m)$ as an $(N-1)$-form);
the last condition expresses closedness. 
The cocycle proportional to $c_1$ was discovered by 
Rao and Moody \cite{RM94}, and the one proportional to $c_2$ by
this author \cite{Lar91}. 

There is also a similar multi-dimensional generalization of affine
Kac-Moody algebras, sometimes called the central extension. This
term is somewhat misleading because the extension does not commute with
diffeomorphisms, although it does commute with all gauge
transformations.
Let $\oj$ be a finite-dimensional Lie algebra with structure
constants $f_{ab}{}^c$ and Killing metric $\delta_{ab}$. The central
extension of the current algebra $\map(N,\oj)$ is defined by the
brackets
\bes
[J_a(m), J_b(n)] &=& f_{ab}{}^c J_c(m+n)
  + k \delta_{ab} m_\rho S^\rho(m+n), \nl
{[}J_a(m), S^\mu(n)] &=& [S^\mu(m), S^\nu(n)] = 0, 
\label{affine}\\
m_\mu S^\mu(m) &\equiv& 0.
\eens
This algebra admits an intertwining action of the $N$-dimensional
Virasoro algebra (\ref{mVir}):
\be
[L_\mu(m), J_a(n)] = n_\mu J_a(m+n).
\ee

The current algebra $\map(N,\oj)$ also admits another type of
extension in some dimensions. The best known example is the
Mickelsson-Faddeev algebra, relevant for the conventional
anomalies in field theory, which arise when chiral fermions are
coupled to gauge fields in three spatial dimensions. Let
$d_{abc} = \tr \{J_a, J_b\}J_c$ be the totally symmetric third
Casimir operator, and let $\eps^{\mu\nu\rho}$ be the totally
anti-symmetric epsilon tensor in three dimensions. The 
Mickelsson-Faddeev algebra \cite{Mi89} reads in a Fourier basis:
\bes
[J_a(m), J_b(n)] &=& f_{ab}{}^c J_c(m+n)
  + d_{abc} \epsilon^{\mu\nu\rho} m_\mu n_\nu A^c_\rho(m+n), \nl
{[}J_a(m), A^b_\nu(n)] &=& -f_{ac}{}^b A^c_\nu(m+n) 
  + \delta_a^b m_\nu \dlt(m+n), 
\label{MF}\\
{[}A^a_\mu(m), A^b_\nu(n)] &=& 0.
\eens
$A^a_\mu(m)$ are the Fourier components of the gauge connection.

Note that $Q_a \equiv J_a(0)$ generates a Lie algebra isomorphic to
$\oj$, whose Cartan subalgebra is identified with the charges.
Moreover, the subalgebra of (\ref{affine}) spanned by 
$J_a(m_0) \equiv J_a(m)$, where $m = (m_0, 0, ..., 0) \in \ZZ$, reads
\be
[J_a(m_0), J_b(n_0)] &=& f_{ab}{}^c J_c(m_0+n_0)
+ k \delta_{ab} m_0 \dlt(m_0+n_0),
\ee
where $k = S^0(0)$,
which we recognize as the affine algebra $\widehat{\oj}$. Since all
non-trivial unitary irreps of $\widehat{\oj}$ has $k>0$ \cite{GO86}, 
it is impossible to combine unitary and non-zero $\oj$ charges also
for the higher-dimensional algebra (\ref{affine}).
This follows immediately from the fact that the restriction of a 
unitary irrep to a subalgebra is also unitary (albeit in general
reducible).

In contrast, the Mickelsson-Faddeev algebra (\ref{MF}) has apparently no
faithful unitary representations on a separable Hilbert space
\cite{Pic89}. A simple way to understand this is to note that the
restriction to every loop subalgebra is proper and hence lacks unitary
representations of lowest-weight type \cite{Lar05b}. This presumably means
that this kind of extension should be avoided. Indeed, Nature appears to
abhor this kind of anomaly, which is proportional to the third Casimir.

\section{DGRO algebra}
\label{sec:DGRO}

The Fourier formalism in the previous section makes the analogy with
the usual Virasoro algebra manifest, but it is neither illuminating
nor a useful starting point for representation theory. To bring out
the geometrical content, we introduce the DGRO 
{\em (Diffeomorphism, Gauge, Repara\-metri\-zation, Observer)} algebra
$DGRO(N,\oj)$, whose ingredients are spacetime diffeomorphisms which 
generate $\vect(N)$,
repara\-metri\-zations of the observer's trajectory which form an
additional $\vect(1)$ algebra, and gauge transformations which generate
a current algebra. Classically, the algebra is $\dmap\oplus\vect(1)$.

Let $\xi=\xmu(x)\dmu$, $x\in\RR^N$, $\dmu = \d/\d x^\mu$,
be a vector field, with commutator 
$[\xi,\eta] \equiv \xmu\dmu\ynu\dnu - \ynu\dnu\xmu\dmu$,
and greek indices $\mu,\nu = 1,2,..,N$ label the 
spacetime coordinates. 
The Lie derivatives $\L_\xi$ are the generators of $\vect(N)$.

Let $f = f(t)d/dt$, $t\in S^1$, be a vector field in one dimension. 
The commutator reads 
$[f,g] = (f\dot g - g\dot f)d/dt$, where the dot denotes the $t$ 
derivative: $\dot f \equiv df/dt$. We will also use $\d_t = \d/\d t$
for the partial $t$ derivative.
The choice that $t$ lies on the circle is physically unnatural and is
made for technical simplicity only (quantities can be expanded in
Fourier series). However, this seems to be a
minor problem at the present level of understanding.
Denote the repara\-metri\-zation generators $L_f$.

Let $\map(N,\oj)$ be the current algebra corresponding to the
finite-dimen\-sional semisimple Lie algebra $\oj$ with basis $J_a$, 
structure constants $f_{ab}{}^c$, and Killing metric $\dlt_{ab}$. 
The brackets in $\oj$ are given by (\ref{oj}).
A basis for $\map(N,\oj)$ is given by $\oj$-valued functions $X=X^a(x)J_a$ 
with commutator $[X,Y]=f_{ab}{}^c X^aY^bJ_c$. The intertwining
$\vect(N)$ action is given by $\xi X = \xmu\dmu X^a J_a$. Denote the
$\map(N,\oj)$ generators by $\J_X$.

Finally, let $Obs(N)$ be the space of local functionals of the
observer's trajectory $q^\mu(t)$, i.e. polynomial functions of  
$q^\mu(t)$, $\dot q^\mu(t)$, ... $d^k  q^\mu(t)/dt^k$,  $k$ finite, 
regarded as a commutative algebra. $Obs(N)$ is a $\vect(N)$ module in a 
natural manner.

$DGRO(N,\oj)$ is an abelian but non-central Lie algebra extension of 
$\vect(N) \ltimes \map(N,\oj) \oplus \vect(1)$ by $Obs(N)$:
\[
0 \longrightarrow Obs(N) \longrightarrow DGRO(N,\oj) \longrightarrow
 \vect(N)\ltimes \map(N,\oj)\oplus \vect(1) \longrightarrow 0.
\]
The brackets are given by
\bes
[\L_\xi,\L_\eta] &=& \L_{[\xi,\eta]} 
 + {1\/2\pi i}\int dt\ \dot q^\rho(t) 
 \Big\{ c_1 \d_\rho\dnu\xmu(q(t))\dmu\ynu(q(t)) +\nl
&&\quad+ c_2 \d_\rho\dmu\xmu(q(t))\dnu\ynu(q(t)) \Big\}, \nl
{[}\L_\xi, \J_X] &=& \J_{\xi X}, \nl
{[}\J_X, \J_Y] &=& \J_{[X,Y]} - {c_5\/2\pi i}\dlt_{ab}
 \int dt\ \dot q^\rho(t)\d_\rho X^a(q(t))Y^b(q(t)), \nl
{[}L_f, \L_\xi] &=& {c_3\/4\pi i} \int dt\ 
 (\ddot f(t) - i\dot f(t))\dmu\xmu(q(t)), 
\label{DGRO}\\
{[}L_f,\J_X] &=& 0, \nl
{[}L_f,L_g] &=& L_{[f,g]} 
 + {c_4\/24\pi i}\int dt (\ddot f(t) \dot g(t) - \dot f(t) g(t)), \nl
{[}\L_\xi, q^\mu(t)] &=& \xmu(q(t)), \nl
{[}L_f, q^\mu(t)] &=& -f(t)\dot q^\mu(t), \nl
{[}\J_X,  q^\mu(t)] &=& {[} q^\mu(s),  q^\nu(t)] = 0,
\eens
extended to all of $Obs(N)$ by Leibniz' rule and linearity.
The numbers $c_1-c_5$ are called {\em abelian charges}, in analogy with
the central charge of the Virasoro algebra. In \cite{Lar98,Lar01}
slightly more complicated extensions were considered, which
depend on three additional abelian charges $c_6-c_8$. However, these
vanish automatically when $\oj$ is semisimple.

\section{Representations of the DGRO algebra}
\label{sec:DGROrep}

To construct Fock representations of the ordinary Virasoro algebra is
straightforward:
\begin{itemize}
\item
Start from classical modules, i.e. primary fields = scalar densities.
\item
Introduce canonical momenta.
\item
Normal order.
\end{itemize}
The first two steps of this procedure generalize nicely to higher
dimensions. The classical representations of the DGRO algebra are tensor
fields over $\RR^N \times S^1$ valued in $\oj$ modules. The basis of a
classical DGRO module $\QQ$ is thus a field
$\fa(x,t)$, $x\in\RR^N$, $t\in S^1$, 
where $\al$ is a collection of all kinds of indices.
The $DGRO(N,\oj)$ action on $\QQ$ can be succinctly summarized as
\bes
[\L_\xi, \fa(x,t)] &=& -\xmu(x)\dmu\fa(x,t)
- \dnu\xmu(x)T^{\al\nu}_{\bt\mu}\fb(x,t), \nl
{[}\J_X, \fa(x,t)] &=& -X^a(x)J^{\al}_{\bt a}\fb(x,t), 
\\
{[}L_f, \fa(x,t)] &=& -f(t)\d_t\fa(x,t) 
- \la(\dot f(t)-if(t))\fa(x,t).
\eens
Here $J_a = (J^\al_{\bt a})$ and $T^\mu_\nu = (T^{\al\mu}_{\bt\nu})$ are
matrices satisfying $\oj$ (\ref{oj}) and $gl(N)$, respectively:
\be
[T^\mu_\nu,T^\si_\tau] = 
\dlt^\si_\nu T^\mu_\tau - \dlt^\mu_\tau T^\si_\mu.
\label{glN}
\ee
The tensor field representations of the DGRO algebra can thus be
expressed in matrix form as
\bes
\L_\xi &=& -\int \dNx\int dt\ 
(\xmu(x)\dmu\fa(x,t) +\dnu\xmu(x)T^{\al\nu}_{\bt\mu}\fb(x,t))
\pa(x,t), \nl
\J_X &=& -\int \dNx\int dt\ 
X^a(x)J^{\al}_{\bt a}\fb(x,t) \pa(x,t), 
\label{class} \\
L_f &=& -\int \dNx \int dt\ 
(f(t)\d_t\fa(x,t) - \la(\dot f(t)-if(t))\fa(x,t)) \pa(x,t),
\eens
where the conjugate momentum $\pa(x,t) = \dd{\fa(x,t)}$ satisfies
\be
[\pa(x,t), \fb(x',t')] = \dlt^\bt_\al \dlt(x-x') \dlt(t-t').
\ee

However, the normal-ordering step simply does not work in several 
dimensions, because
\begin{itemize}
\item
It requires that a foliation of spacetime into space and time has been 
introduced, which runs against the idea of diffeomorphism invariance.
\item
Normal ordering of bilinear expressions always results in a {\em central}
extension, but the Virasoro cocycle is non-central when $N\geq 2$.
\item
It is ill defined. Formally, attempts to normal order result in an
{\em infinite} central extension, which of course makes no sense.
\end{itemize}
To avoid this problem,
the crucial idea in \cite{Lar98} was to expand all fields in a Taylor 
series around the observer's trajectory and truncate at order $p$,
before introducing canonical momenta. 
Hence we expand e.g.,
\be
\fa(x,t) = \summ p {1\/\mm!} \fam(t)(x-q(t))^\mm,
\label{Taylor}
\ee
where $\mm = (m_1, \ab m_2, \ab ..., \ab m_N)$, all $m_\mu\geq0$, is a 
multi-index of length $|\mm| = \sum_{\mu=1}^N m_\mu$,
$\mm! = m_1!m_2!...m_N!$, and
\be
(x-q(t))^\mm = (x^1-q^1(t))^{m_1} (x^2-q^2(t))^{m_2} ...
 (x^N-q^N(t))^{m_N}.
\label{power}
\ee
Denote by $\mu$ a unit vector in the $\mu$:th direction, so that
$\mm+\mu = (m_1, \ab ...,m_\mu+1, \ab ..., \ab m_N)$, and let
\be
\fam(t) = \d_\mm\fa(q(t),t)
= \underbrace{\d_1 .. \d_1}_{m_1} .. 
\underbrace{\d_N .. \d_N}_{m_N} \fa(q(t),t)
\label{jetdef}
\ee
be an $|\mm|$:th order mixed partial derivative of $\fa(x,t)$ evaluated 
on the observer's trajectory $q^\mu(t)$. 

Given two jets $\phi_\cmm(t)$ and $\psi_\cmm(t')$, we define their product
\be
(\phi(t)\psi(t'))_\cmm 
= \sum_\nn {\mm\choose\nn}\phi_\cnn(t) \psi_{,\mm-\nn}(t'),
\label{product}
\ee
where
\be
{\mm\choose\nn} = {\mm!\/\nn!(\mm-\nn)!} = 
{m_1\choose n_1}{m_2\choose n_2}...{m_N\choose n_N}.
\ee
It is clear that $(\phi(t)\psi(t'))_\cmm$ is the jet corresponding to the
field $\phi(x,t)\psi(x,t')$. For brevity, we also denote
$(\phi\psi)_\cmm(t) = (\phi(t)\psi(t))_\cmm$.

$p$-jets transform under $DGRO(N,\oj)$ as
\bes
[\L_\xi, \fam(t)] &=& \d_\mm([\L_\xi,\fa(q(t),t)]) 
+ [\L_\xi, q^\mu(t)]\dmu\d_\mm\fa(q(t),t) \nl
&\equiv& -\sumnmp{} T^{\al\nn}_{\bt\mm}(\xi(q(t))) \fbn(t), \nl
{[}\J_X, \fam(t)] &=& \d_\mm([\J_X,\fa(q(t),t)]) 
\label{jet} \\
&\equiv& -\sumnmp{} J^{\al\nn}_{\bt\mm}(X(q(t))) \fbn(t), \nl
{[}L_f, \fam(t)] &=& -f(t)\dot\fam(t) 
- \la(\dot f(t)-if(t))\fam(t),
\eens
where
\bes
T^\mm_\nn(\xi) &\equiv& (T^{\al\mm}_{\bt\nn}(\xi))
= \sum_{\mu\nu}{\nn\choose\mm} \d_{\nn-\mm+\nu}\xmu T^\nu_\mu \nl
&&\qquad + \sum_\mu{\nn\choose\mm-\mu}\d_{\nn-\mm+\mu}\xmu
  - \sum_\mu \dlt^{\mm-\mu}_\nn \xmu,
\label{Tmn}\\
J^\mm_\nn(X) &\equiv& (J^{\al\mm}_{\bt\nn}(X))
= {\nn\choose\mm} \d_{\nn-\mm} X^a J_a.
\eens

We thus obtain a non-linear realization of $\vect(N)$ on the space of
trajectories in the space of tensor-valued
$p$-jets\footnote{$p$-jets are usually defined
as an equivalence class of functions: two functions are equivalent if all
derivatives up to order $p$, evaluated at $q^\mu$, agree. However, each
class has a unique representative which is a polynomial of order at most
$p$, namely the Taylor expansion around $q^\mu$, so we may canonically
identify jets with truncated Taylor series. Since $q^\mu(t)$ depends on a
parameter $t$, we deal in fact with trajectories in jet space, but these
will also be called jets for brevity.}; 
denote this space by $J^p\QQ$. Note that $J^p\QQ$ is spanned by $q^\mu(t)$
and $\{\fam(t)\}_{|\mm|\leq p}$ and thus it is not a $DGRO(N,\oj)$ module 
by itself, because diffeomorphisms act non-linearly on $q^\mu(t)$, as can
be seen in (\ref{DGRO}). However, the space $C(J^p\QQ)$ of functionals on
$J^p\QQ$ (local in $t$) {\em is} a module, because the action on a $p$-jet
can never produce a jet of order higher than $p$. The space
$C(q) \otimes_q J^p\QQ$, where only the trajectory itself appears 
non-linearly, is a submodule.

The crucial observation is that the jet space $J^p\QQ$
consists of finitely many functions of a single
variable $t$, which is precisely the situation where the normal ordering
prescription works. After normal ordering, denoted by double dots $:\ :$,
we obtain a Fock representation of the DGRO algebra:
\bes
\L_\xi &=& \int dt\ \Big\{ \no{\xmu(q(t))  p_\mu(t)} -
\sum_{|\nn|\leq|\mm|\leq p}
T^{\al\nn}_{\bt\mm}(\xi(q(t))) \no{ \fbn(t)\pam(t) } \Big\}, \nl
\J_X &=& -\int dt\ \Big\{ \sum_{|\nn|\leq|\mm|\leq p}
J^{\al\nn}_{\bt\mm}(\xi(q(t))) \no{ \fbn(t)\pam(t) } \Big\}, 
\label{Fock}\\
L_f &=& \int dt\ \Big\{ -f(t)\no{\dot\fam(t)\pam(t)} 
- \la(\dot f(t)-if(t))\no{\fam(t)\pam(t)} \Big\},
\eens
where we have introduced canonical momenta
$p_\mu(t) = \dd{ q^\mu(t)}$ and $\pam(t) = \dd{\fam(t)}$.
The field $\fa(x,t)$ can be either bosonic or fermionic but the
trajectory $q^\mu(t)$ is of course always bosonic.

Normal ordering is defined with respect to frequency; any function of
$t \in S^1$ can be expanded in a Fourier series, e.g.
\be
 p_\mu(t) = \intdm \hat p_\mu(m) \e^{-imt} \equiv
 p_\mu^<(t) + \hat p_\mu(0) +  p_\mu^>(t),
\label{Fourier}
\ee
where $p_\mu^<(t)$ ($p_\mu^>(t)$) is the sum over negative (positive) 
frequency modes only. Then
\be
\no{\xmu(q(t))  p_\mu(t)} 
\equiv \xmu(q(t))  p_\mu^<(t) +  p_\mu^>(t) \xmu(q(t)),
\ee
where the zero mode has been included in $p_\mu^<(t)$.

It is clear that (\ref{Fock}) defines a Fock representation for every
$gl(N)$ irrep $\rep$ and every $\oj$ irrep $M$; denote this Fock space
by $J^p\FF$, which indicates that it also depends on the truncation 
order $p$. Namely, introduce a
Fock vacuum $\ket0$ which is annihilated by half of the oscillators,
i.e.
\be
\phi^{\alpha<}_\cmm(t)\ket 0 = \pi_{\alpha<}^\cmm(t)\ket 0 =  
q^\mu_<(t)\ket 0 = p_\mu^<(t)\ket0 = 0.
\label{LER}
\ee
Then $DGRO(N,\oj)$ acts on the space of functionals 
$C( q^\mu_>, p_\mu^>,\phi^{\alpha>}_\cmm,\pi_{\alpha>}^\cmm)$ 
of the remaining oscillators; this is the Fock module.
Define numbers $k_0(\rep)$, $k_1(\rep)$, $k_2(\rep)$ and $y_M$ by
\bes
\trrep T^\mu_\nu &=& k_0(\rep) \dlt^\mu_\nu, \nl
\trrep T^\mu_\nu T^\si_\tau &=&
 k_1(\rep) \dlt^\mu_\tau \dlt^\si_\nu 
 + k_2(\rep) \dlt^\mu_\nu \dlt^\si_\tau, \\
\trM J_aJ_b &=& y_M \dlt_{ab}.
\eens
For an unconstrained tensor with $p$ upper and $q$ lower indices and
weight $\ka$, we have
\bes
\dim(\rep) = N^{p+q}, &\quad&
k_0(\rep)= -(p-q-\ka N) N^{p+q-1},
\label{krep}\\
k_1(\rep) = (p+q)N^{p+q-1}, &\quad&
k_2(\rep) = ((p-q-\ka N)^2 - p - q) N^{p+q-2}.
\eens
Note that if $\ka = (p-q)/N$, $\rep$ is an $sl(N)$ representation.
For the symmetric representations on $\ell$ lower indices, $S_\ell$,
and on $\ell$ upper indices, $S^\ell$, we have
\bes
\dim(S_\ell) = \dim(S^\ell) &=& {N-1+\ell\choose\ell}, \nl
k_0(S_\ell) = -k_0(S^\ell) &=& {N-1+\ell\choose\ell-1},
\nlb{kSl}
k_1(S_\ell) = k_1(S^\ell) &=& {N+\ell\choose\ell-1}, \nl
k_2(S_\ell) = k_2(S^\ell) &=& {N-1+\ell\choose\ell-2}.
\eens

The values of the abelian charges $c_1 - c_5$ (\ref{DGRO}) 
were calculated in \cite{Lar98}, Theorems 1 and 3, and in 
\cite{Lar01}, Theorem 1:
\bes
c_1 &=& 1 - u\Np{} - x {N+p+1\choose N+2}, \nl
c_2 &=& -v\Np{} - 2w \Np{+1} - x {N+p\choose N+2}, \nl
c_3 &=& 1 + (1-2\la) ( w\Np{} + x\Np{+1}), 
\label{cs}\\
c_4 &=& 2N - x(1-6\la+6\la^2)\Np{}, \nl
c_5 &=& y\Np{}.
\eens
where 
\bes
u = \mp k_1(\rep)\, \dim\,M, &\qquad&
x = \mp \dim\,\rep\,\dim\,M, \nl
v = \mp k_2(\rep)\, \dim\,M, &\qquad&
y = \mp \dim\,\rep\,y_M, 
\label{numdef}\\
w = \mp k_0(\rep)\, \dim\,M, &&
\eens
and the sign factor depends on the Grassmann parity of $\fa$; the upper
sign holds for bosons and the lower for fermions, respectively.
The $p$-independent contributions to $c_1$, $c_3$ and $c_4$ come from
the trajectory $q^\mu(t)$ itself.

\section{MCCQ: Manifestly Covariant Canonical Quantization}
\label{sec:MCCQ}

Consider a classical dynamical system with action $S$ and degrees of
freedom $\fa$. As is customary in the antifield literature \cite{HT92},
we use an abbreviated notation where the index $\alpha$ stands for both
discrete indices and spacetime coordinates.
Dynamics is governed by the Euler-Lagrange (EL) equations,
\be
\Ea = \da S \equiv {\dlt S\/\dlt\fa} = 0.
\label{EL}
\ee
We do not assume that the EL equations are all independent; rather,
let there be identities of the form
\be
r^\al_a\Ea \equiv 0,
\label{Rident}
\ee
where the $r^\al_a$ are some functionals of $\fa$. 

Introduce an antifield $\fsa$ for each EL equation (\ref{EL}), and a
second-order antifield $\za$ for each identity (\ref{Rident}).
Replace the space of $\phi$-histories $\QQ$ by the extended history
space $\QQ^*$, spanned by both $\fa$, $\fsa$, and $\za$.
In $\QQ^*$ we define the Koszul-Tate (KT) differential $\dlt$ by
\bes
\dlt \fa &=& 0, \nl
\dlt \fsa &=& \Ea, 
\label{dlt0}\\
\dlt \za &=& r^\al_a\fsa.
\eens
One checks that $\dlt$ is nilpotent, $\dlt^2 = 0$. Define the antifield
number $\afn \fa = 0$, $\afn \fsa = 1$, $\afn \za = 2$. The KT
differential clearly has antifield number $\afn\dlt = -1$.

The space $C(\QQ^*)$ decomposes into subspaces $C^k(\QQ^*)$ of
fixed antifield number
\be
C(\QQ^*) = \sum_{k=0}^\infty C^k(\QQ^*)
\ee
The KT complex is 
\be
0 \larroww \dlt C^0 \larroww \dlt C^1 \larroww \dlt 
C^2 \larroww \dlt \ldots
\label{complex1}
\ee
The cohomology spaces are defined as usual by
$H_\cl^\bullet(\dlt) = \ker\dlt/\im\dlt$, i.e.
$H_\cl^k(\dlt) = (\ker\dlt)_k/(\im\dlt)_k$, where
the subscript $\cl$ indicates that we deal with a classical phase space.
It is easy to see that
\bes
(\ker \dlt)_0 &=& C(\QQ), \nle
(\im \dlt)_0 &=& C(\QQ)\Ea \equiv \N.
\eens
Thus $H_\cl^0(\dlt) = C(\QQ)/\N = C(\Sigma)$.
The higher cohomology groups vanish, because although $r^\al_a\fsa$ is
KT closed ($\dlt(r^\al_a\fsa) = r^\al_a\Ea \equiv 0$), it is also KT
exact ($r^\al_a\fsa = \dlt \za$). The complex 
(\ref{complex1}) thus gives us
a resolution of the covariant phase space $C(\Sigma)$, which by
definition means that
$H_\cl^0(\dlt) = C(\Sigma)$, $H_\cl^k(\dlt) = 0$, for all $k>0$.

Alas, the antifield formalism is not suited for canonical quantization.
We can define an antibracket in
$\QQ^*$, but in order to do canonical quantization we need an honest
Poisson bracket. To this end, we introduce canonical momenta conjugate to
the history and its antifields, and obtain an even larger space $\PP^*$,
which may be thought of as the phase space corresponding to the 
extended history space $\QQ^*$.
Introduce canonical momenta $\pa = \dd{\fa}$, $\psa = \dd{\fsa}$ and 
$\chi^a = \dd{\za}$, which satisfy the graded canonical commutation 
relations ($\fa$ is assumed bosonic),
\be
[\pb,\fa] = \dlt^\al_\bt, \quad
\{\psb,\fsa\} = \dlt_\al^\bt, \quad
[\chi^a, \zb] = \dlt^a_b,
\label{ccr*}
\ee
where $\{\cdot,\cdot\}$ is the symmetric bracket.
Let $\PP$ be the phase space of histories with basis $(\fa,\pb)$,
and let $\PP^*$ be the extended phase space with basis
$(\fa,\pb,\fsa,\psb,\ab\za,\chi^b)$. 
The definition of the KT differential extends to $\PP^*$ by 
requiring that $\dlt F = [Q_{KT},F]$ for every $F \in C(\PP^*)$, 
where the nilpotent KT operator is
\be
Q_{KT} =  \Ea\psa + r^\al_a\fsa\chi^a.
\label{KT}
\ee
{F}rom this explicit expression we can read off the action of $\dlt$
on the canonical momenta. As explained in \cite{Lar04}, the momenta
are killed in cohomology, provided that the Hessian 
$\dlt^2 S/\dlt\fa\dlt\fb$ is non-singular.

The identity (\ref{Rident})
implies that $J_a = r^\al_a\pa$ generate a Lie algebra under the Poisson
bracket. Namely, all $J_a$'s preserve the action, because
\be
[J_a, S] = r^\al_a[\pa,S] = r^\al_a\Ea \equiv 0,
\ee
and the bracket of two operators which preserve some structure also 
preserves the same structure.
We will only consider the case that the $J_a$'s generate a proper Lie
algebra $\oj$ as in (\ref{oj}).
The formalism extends without too much extra work to the more general
case of structure functions $f_{ab}{}^c(\phi)$, but we will not need
this complication here.
It follows that the functions $r^\al_a$ satisfy the identity
\be
\db r^\al_b r^\bt_a - \db r^\al_a r^\bt_b = f_{ab}{}^c r^\al_c.
\label{rident}
\ee
The Lie algebra $\oj$ also acts on the antifields:
\bes
[J_a, \fa] &=& r^\al_a, \nl
{[}J_a, \fsa] &=& -\da r^\bt_a \fsb
\label{Jf}\\
{[}J_a, \zb] &=& f_{ab}{}^c\zc.
\eens
In particular, it follows that $\fsa$ carries a $\oj$ representation 
because it transforms in the same way as $\pa$ does.

Classically, it is always possible to reduce the phase space further,
by identifying points on $\oj$ orbits. To implement this additional
reduction, we introduce ghosts $c^a$ with anti-field number
$\afn c^a = -1$, and ghost momenta $b_a$ satisfying
$\{b_a, c^b\} = \dlt^b_a$.
The Lie algebra $\oj$ acts on the ghosts as 
$[J_a, c^b] = -f_{ac}{}^b c^c$.
The full extended phase space, still denoted by $\PP^*$, is spanned by
$(\fa,\pb,\ab \fsa,\psb,\ab \za,\chi^b,\ab c^a,b_b)$.
The generators of $\oj$ are thus identified with the following vector
fields in $\PP^*$:
\bes
J_a &=& r^\al_a\pa -\da r^\bt_a\fsb\psa + f_{ab}{}^c\zc\chi^b
- f_{ab}{}^c c^b b_c 
\nlb{Ja}
&=& J^{field}_a + J^{ghost}_a,
\eens
where $J^{ghost}_a = -f_{ab}{}^c c^b b_c$ and $J^{field}_a$ is the rest.

Now define the longitudinal derivative $d$ by
\bes
d c^a &=& -\half f_{bc}{}^a c^b c^c, \nl
d \fa &=& r^\al_a c^a, \nle
d \fsa &=& \da r^\bt_a \fsb c^a, \nl
d \za &=& - f_{ab}{}^c \zc c^b.
\eens
The longitudinal derivative can be written as $dF = [Q_{Long},F]$
for every $F\in C(\PP^*)$, where 
\be
Q_{Long} = J^{field}_a c^a - \half f_{ab}{}^c c^a c^b b_c 
= J^{field}_a c^a + \half J^{ghost}_a c^a.
\label{QLong1}
\ee
We note that $Q_{Long}$ can be considered as smeared gauge generators,
$\J_X = X^a J_a$, where the smearing function $X^a$ is the fermonic
ghost $c^a$:
\be
Q_{Long} = \J^{field}_c + \half \J^{ghost}_c.
\label{smear}
\ee

One verifies that $d^2 = 0$ when acting on the fields and antifields by
means of the identify (\ref{rident}) and the Jacobi identities for $\oj$.
Moreover, it is straightforward to show that $d$ anticommutes with the
KT differential, $d\dlt = -\dlt d$; the proof is again done by checking
the action on the fields.
Hence we may define the nilpotent {\em BRST derivative} $s = \dlt+d$,
\bes
s c^a &=& -\half f_{bc}{}^a c^b c^c, \nl
s \fa &=& r^\al_a c^a, 
\nlb{sfields}
s \fsa &=& \Ea + \da r^\bt_a \fsb c^a, \nl
s \za &=& r^\al_a\fsa - f_{ab}{}^c \zc c^b.
\eens
Nilpotency immediately follows because 
$s^2 = \dlt^2 + \dlt d + d\dlt + d^2 = 0$.
The BRST operator can be written in the form $sF = [Q_{BRST},F]$ with
\bes
Q_{BRST} &=& Q_{KT} + Q_{Long} \nl
&=&  \Ea\psa + r^\al_a\fsa\chi^a + J^{field}_a c^a 
+ \half J^{ghost}_a c^a \nle
&=&  -\half f_{ab}{}^c c^a c^b b_c + r^\al_a c^a \pa
+ (\Ea + \da r^\bt_a \fsb c^a) \psa \nl
&&+ (r^\al_a\fsa - f_{ab}{}^c \zc c^b) \chi^a.
\eens

\section{MCCQ: Jets and covariant quantization}
\label{sec:MCCQ-jets}

Like $C(\QQ^*)$, the space $C(\PP^*)$ of functions over the extended 
history phase space decomposes into subspaces $C^k(\PP^*)$ of fixed 
antifield number,
\be
C(\PP^*) = \sum_{k=-\infty}^\infty C^k(\PP^*)
\ee
The complexes associated with $Q_{BRST}$, $Q_{KT}$, and $Q_{Long}$ now
extend to infinity in both directions:
\be
\ldots \larroww Q C^{-2} \larroww Q C^{-1} \larroww
0 \larroww Q C^0 \larroww Q C^1 \larroww Q C^2 \larroww Q \ldots
\label{complex2}                      
\ee
It is important that the spaces $C^k(\PP^*)$ are phase spaces, equipped
with the Poisson bracket (\ref{ccr*}). Unlike the resolution
(\ref{complex1}), the new resolution (\ref{complex2}) therefore allows
us to do canonical quantization: replace Poisson brackets by commutators
and represent the graded Heisenberg algebra (\ref{ccr*}) on a Hilbert
space. However, the Heisenberg algebra can be represented on different
Hilbert spaces; there is no Stone-von Neumann theorem in infinite
dimension. To pick the correct one, we must impose the physical
condition that there is an energy which is bounded on below.

To define the Hamiltonian, we must single out a privileged variable 
$t$ among the $\al$'s, and declare it to be time. However, this amounts
to an a-priori foliation of spacetime into space and time, which clashes 
with the philosophy
of diffeomorphism invariance. Therefore, we now introduce the observer's
trajectory $q^\mu(t)$, which locally defines a time direction (namely
$\dot q^\mu(t)$). 
Thus, we reformulate the classical theory in jet coordinates.
To the fields $c^a(x)$, $\fa(x)$, $\fsa(x)$ and $\za(x)$ we associate
$p$-jets $c^a_\cmm(t)$, $\fam(t)$, $\fsam(t)$ and $\zam(t)$, as in
(\ref{jetdef}). The jets have
canonical momenta $b_a^\cmm(t) = \dd{c^a_\cmm(t)}$,
$\pam(t) = \dd{\fam(t)}$, $\psam(t) = \dd{\fsam(t)}$ and 
$\chi^{a\cmm}(t) = \dd{\zam(t)}$, defined as usual by commutation
relations
\bes
[\pbn(t),\fam(t')] &=& \dlt^\al_\bt \dlt^\nn_\mm \dlt(t-t'), \nle
\{\psbn(t),\fsam(t')\} &=& \dlt_\al^\bt \dlt^\nn_\mm \dlt(t-t'),
\eens
etc. Note that the jet momenta $\pam(t)$ are {\em not} the Taylor
coefficients of the momentum $\pa(x)$; a jet, defined by
(\ref{Taylor}), always has a multi-index downstairs.

The jets depend on an additional parameter $t$, and a Taylor series
like (\ref{Taylor}) defines a field $\fa(x,t)$ rather than $\fa(x)$.
But the fields in the physical phase space do not depend on the parameter
$t$, wherefore we must impose extra conditions, e.g.
\be
\d_t\fa(x,t) \equiv {\d\fa(x,t)\/\d t} = 0.
\ee
We can implement this condition by introducing new antifields 
$\bar\fa(x,t)$.
However, the identities $\d_t\Ea(x,t) \equiv 0$ give rise to unwanted
cohomology. To kill this condition, we must introduce yet another
antifield $\bar\fsa(x,t)$, etc., as described in detail in \cite{Lar04}.
This means that our extended jet space $\QQ^*$ is spanned by the
jets 
\[
c^a_\cmm(t), \fam(t), \fsam(t), \zam(t), 
\bar c^a_\cmm(t), \bar \fam(t), \bar \fsam(t), \bar \zam(t),
\]
and that the extended jet phase space $\PP^*$ in addition has basis
vectors
\[
b_a^\cmm(t), \pam(t), \psam(t), \chi^{a\cmm}(t), 
\bar b_a^\cmm(t), \bar \pam(t), \bar \psam(t), \bar \chi^{a\cmm}(t).
\]
The cohomology is not changed by the new, barred, variables, since all we
have done is to introduce an extra variable $t$ and then eliminate it
in cohomology.

The Taylor expansion requires that we introduce the observer's
trajectory as a physical field, but what equation
of motion does it obey? The obvious answer is the geodesic equation,
which we compactly write as $\GG_\mu(t) = 0$. The geodesic operator
$\GG_\mu(t)$ is a function of the metric $g_{\mu\nu}(q(t),t)$
and its derivatives on the curve $q^\mu(t)$. To eliminate this ideal
in cohomology we introduce the trajectory antifield $\qsmu(t)$, and
extend the KT differential to it:
\bes
\dlt  q^\mu(t) &=& 0, 
\nlb{geo}
\dlt \qsmu(t) &=& \GG_\mu(t).
\eens
For models defined over Minkowski spacetime, the geodesic equation
simply becomes $\ddot q^\mu(t) = 0$, and the KT differential reads
\be
\dlt \qsmu(t) = \eta_{\mu\nu} q^\nu(t).
\ee
$H_\cl^0(\dlt)$ only contains trajectories which are straight lines,
\be
 q^\mu(t) = u^\mu t + a^\mu,
\label{line}
\ee
where $u^\mu$ and $a^\mu$ are constant vectors. We may also require
that $u^\mu$ has unit length, $u_\mu u^\mu = 1$. This condition fixes
the scale of the parameter $t$ in terms of the Minkowski metric, so
we may regard it as proper time rather than as an arbitrary parameter.
We define momenta $p_\mu(t) = \dd{q^\mu(t)}$ and 
$\psmu(t) = \dd{\qsmu(t)}$ for the observer's trajectory 
and its antifield. 

We can now define a
genuine Hamiltonian $H$, which translates the fields relative to the
observer or vice versa. Since the formulas are shortest when $H$ acts
on the trajectory but not on the jets, we make that choice, and define
\be
H = i\int dt\ (\dot q^\mu(t) p_\mu(t) + \dot\qsmu(t)\psmu(t)).
\label{H2}
\ee
Note the sign; moving the fields forward in $t$ is equivalent to moving
the observer backwards.
{F}rom (\ref{Taylor}) we get the energy of the fields:
\be
[H, \fa(x,t)] = -i\dot q^\mu(t)\dmu\fa(x,t).
\label{Hcl}
\ee
This a crucial result, because it allows us to define a
genuine energy operator in a covariant way. 
In Minkowski space, the trajectory is a straight line (\ref{line}),
and $\dot q^\mu(t) = u^\mu$. If we take $u^\mu$ to be the constant
four-vector $u^\mu = (1,0,0,0)$, then (\ref{Hcl}) reduces to
\be
[H, \fa(x,t)] = -i {\d\/\d x^0}\fa(x,t).
\ee
Equation (\ref{H2}) is thus a genuine covariant generalization of
the energy operator.

Now we quantize the theory.
Since all operators depend on the parameter $t$, we can define
the Fourier components as in (\ref{Fourier}).
The the Fock vacuum $\ket 0$ is defined to be annihilated by
all negative frequency modes, $\fam(-m)$, $q^\mu(-m)$, etc. with $-m<0$.
The normal-ordered form of the Hamiltonian (\ref{H2}) reads, in
Fourier space,
\be
H = -\intdm m( \no{ q^\mu(m) p_\mu(-m)} + \no{\qsmu(m) p_\mu(-m)} ),
\label{Hq}
\ee
where double dots indicate normal ordering with respect to frequency.
This ensures that $H\ket 0 = 0$.
The classical phase space $H_\cl^0(\dlt)$ is thus the the space of
fields $\fa(x)$ which solve $\Ea(x)=0$,
and trajectories $q^\mu(t) = u^\mu t+a^\mu$, where $u^2 = 1$.
After quantization, the fields and trajectories become operators which
act on the physical Hilbert space $\HH = H_\qm^0(Q_{KT})$, which is the space
of functions of the positive-energy modes of the classical phase
space variables.

This construction differs technically from conventional canonical
quantization, but there is also a physical difference. Consider the
state $\ket{\fa(x)} = \fa(x)\ket 0$ which excites one $\phi$ quantum
from the vacuum. The Hamiltonian yields
\bes
H \ket{\fa(x)} &=& -i\dot q^\mu(t)\dmu\fa(x)\ket 0 \nl
&=& -i\ket{\dot q^\mu(t)\dmu\fa(x)} \\
&=& -i\ket{u^\mu\dmu\fa(x)}.
\eens
If $u^\mu$ were a classical variable, the state $\ket{\fa(x)}$ would
be a superposition of energy eigenstates:
\be
H \ket{\fa(x)} = -iu^\mu\dmu\ket{\fa(x)}.
\ee
In particular, let $u^\mu = (1,0,0,0)$ be a unit vector in the $x^0$
direction and $\fa(x) = \exp(ik\cdot x)$ be a plane wave. We then 
define the state $\ket{0;u,a}$ by
\be
 q^\mu(t)\ket{0;u,a} = (u^\mu t + a^\mu)\ket{0;u,a}.
\label{ua}
\ee
Now write $\ket{k; u,a} =\exp(ik\cdot x)\ket{0;u,a}$ for 
the single-quantum energy eigenstate.
\be
H \ket{k; u,a} = k_\mu u^\mu\ket{k;u,a},
\ee
so the eigenvalue of the Hamiltonian is $k_\mu u^\mu = k_0$, as
expected. Moreover, the lowest-energy condition ensures
that only quanta with positive energy will be excited; if 
$k_\mu u^\mu < 0$ then $\ket{k; u,a} = 0$.

\section{MCCQ: Gauge anomalies}
\label{sec:MCCQ-anomalies}

To be concrete, consider the case that the symmetry is the DGRO
algebra (\ref{DGRO}). To each symmetry, we assign ghosts as in the
following table:
\bes
\barr{|l|lllll|}    
\hline
\hbox{} & \hbox{Gen} & \hbox{Smear} 
& \hbox{Ghost} & \hbox{Momentum} & Q_{Long} \\
\hline
\hbox{Diffeomorphisms} & \L_\xi & \xmu(x) 
& c_{diff}^\mu(x,t) & b^{diff}_\mu(x,t) 
& Q^{diff}_{Long} \\
\hbox{Gauge} & \J_X & X^a(x) 
& c_{gauge}^a(x,t) & b^{gauge}_a(x,t) & Q^{gauge}_{Long} \\
\hbox{Reparametrizations} & L_f & f(t) 
&c_{rep}(t) & b^{rep}(t) & Q^{rep}_{Long} \\
\hline
\earr
\eens
The BRST operator is $Q_{BRST} = Q_{Long} + Q_{KT}$, where the
longitudinal operator is given by the prescription (\ref{smear}).
For brevity, we only write down the formulas for the fields $\fa(x,t)$
and the ghosts; the antifields do of course give rise to additional 
terms.
\bes
Q^{diff}_{Long} &=& -\int \dNx \int dt\ \Big\{
(c_{diff}^\mu(x,t)\dmu\fa(x,t) \nl
&& +\dnu c_{diff}^\mu(x,t)T^{\al\nu}_{\bt\mu}\fb(x,t)) \pa(x,t) \nl
&&+ c_{diff}^\mu(x,t)\dmu c_{diff}^\nu(x,t)b^{diff}_\nu(x,t)
\Big\}, \nl
Q^{gauge}_{Long} &=& -\int \dNx\int dt\  \Big\{
c_{gauge}^a(x,t)J^{\al}_{\bt a}\fb(x,t) \pa(x,t) \nl
&&+\half f_{ab}{}^c c_{gauge}^a(x,t) c_{gauge}^b(x,t) b^{gauge}_c(x,t)
\Big\}, 
\label{QLongC}\\
Q^{rep}_{Long} &=& -\int \dNx \int dt\  \Big\{
(c_{rep}(t)\d_t\fa(x,t) \nl
&&+ \la(\dot c_{rep}(t)-ic_{rep}(t))\fa(x,t)) \pa(x,t) \Big\} \nl
&&- \int dt\ c_{rep}(t)\dot c_{rep}(t) b^{rep}(t).
\eens
These formulas assume that the field $\fa(x)$ transforms as a tensor
field. There is an additional term if the field is a connection, but
this terms does not lead to any complications.
After passage to jet space and normal ordering, we use the prescription 
(\ref{smear}) to find the longitudinal derivative, i.e.
$\d_\mm \xmu \to c^\mu_{diff\cmm}$, $\d_\mm X^a \to c^a_{gauge\cmm}$,
and $f \to c_{rep}$:
\bes
Q^{diff}_{Long} &=& \int dt\ \Big\{ 
c^\mu_{diff,\bf0} p_\mu(t) - \sum_{|\nn|\leq|\mm|\leq p}
T^{\al\nn}_{\bt\mm}(c_{diff}(t)) \no{ \fbn(t)\pam(t) } \nl
&&-  \sum_{|\nn|\leq|\mm|\leq p} \no{ T^{\mu\nn}_{\nu\mm}(c_{diff}(t)) 
c^\nu_{diff,\nn}(t) b_\mu^{diff,\mm}(t) }
\Big\}, \nl
Q^{gauge}_{Long} &=& -\int dt\ \Big\{ 
\sum_{|\nn|\leq|\mm|\leq p}
J^{\al\nn}_{\bt\mm}(c_{gauge}(t)) \no{ \fbn(t)\pam(t) } \nl
&&- \half \sum_{|\nn|\leq|\mm|\leq p}
\no{ J^{a\nn}_{b\mm}(c_{gauge}(t)) 
c^b_{gauge,\nn}(t)b^{gauge,\mm}_a(t)}
\Big\}, 
\label{QLongJ}\\
Q^{rep}_{Long} &=& -\int dt\ \Big\{ 
\summ{p} c_{rep}(t)\no{\dot\fam(t)\pam(t)} \nl
&& +\la\summ{p} (\dot c_{rep}(t)-ic_{rep}(t))\no{\fam(t)\pam(t)} \nl
&&+ \no{ c_{rep}(t)\dot c_{rep}(t) b^{rep}(t)}
\Big\}.
\eens
The matrices are given by (cf. (\ref{Tmn}))
\bes
T^\mm_\nn(c_{diff}(t)) &\equiv& (T^{\al\mm}_{\bt\nn}(c_{diff}(t))) 
=\sum_{\mu\nu} {\nn\choose\mm} c^\mu_{diff,\nn-\mm+\nu}(t) T^\nu_\mu \nl
&& + \sum_\mu {\nn\choose\mm-\mu}c^\mu_{diff,\nn-\mm+\mu}(t) 
 -\sum_\mu \dlt^{\mm-\mu}_\nn c^\mu_{diff,{\bf0}}(t), \nl
J^\mm_\nn(c_{gauge}(t)) &\equiv& (J^{\al\mm}_{\bt\nn}(c_{gauge}(t)))
= {\nn\choose\mm} c^a_{gauge,\nn-\mm}(t) J_a.
\ees
$T^{\mu\mm}_{\nu\nn}$ and $J^{a\mm}_{b\nn}$ denote the specializations
of $T^{\al\mm}_{\bt\nn}$ and $J^{\al\mm}_{\bt\nn}$ to the adjoint
representations;
$\sum_\mm T^{\mu\mm}_{\nu\nn}(c)c^\nu_\cmm = (c^\nu c^\mu_{,\nu})_\cnn$ 
and $\sum_\mm J^{a\mm}_{b\nn}(c)c^b_\cmm = (c^a c^b)_\cnn$

The condition for $Q_{Long}^2 = 0$, and thus $Q_{BRST}^2 = 0$, is
that the algebra generated by the normal-ordered gauge generators
is anomaly free. 
However, even if this condition fails, which is the typical situation,
everything is not lost. The KT operator is still nilpotent, and
we can implement dyna\-mics as the KT cohomology in the extended phase space
without ghosts. The physical phase space now grows, because some gauge
degrees of freedom become physical upon quantization.

\section{Gravity}
\label{sec:Gravity}

Finally we are ready to apply the MCCQ formalism to general relativity.
For simplicity we consider only pure gravity. 
The only field is the symmetric metric $g_{\mu\nu}(x)$. The
inverse $g^{\mu\nu}$, the determinant $g = \det (g_{\mu\nu})$,
the Levi-Civit\`a connection 
$\Gamma^\mu_{\nu\rho}$, Riemann's curvature tensor 
$ R^\rho{}_{\si\mu\nu}$, the Ricci tensor $R_{\mu\nu}$, the
scalar curvature $R = g^{\mu\nu}R_{\mu\nu}$ and the Einstein
tensor $G^{\mu\nu} = R^{\mu\nu} - \half g^{\mu\nu} R$
are defined as usual.
The covariant derivative is
\be
\nabla_\mu = \dmu + \Gamma^\rho_{\nu\mu} T^\nu_\rho,
\ee
where $T^\mu_\nu$ are finite-dimensional matrices satisfying $gl(N)$
(\ref{glN}). 

The Einstein action
\be
S_E = {1\/16\pi} \int \dFx\  \sqrt{g(x)} R(x).
\ee
leads to Einstein's equation of motion
\be
G^{\mu\nu}(x) = 0,
\label{Einstein}
\ee
which is subject to the identity
\be
\nabla_\nu G^{\mu\nu}(x) \equiv 0
\label{Econt}
\ee
We introduce a fermionic antifield $g^{\mu\nu}_*(x)$ for (\ref{Einstein}),
a bosonic second-order antifield $\zeta^\mu(x)$ for (\ref{Econt}), and
a ghost $c^\mu_{diff}(x)$ to eliminate diffeomorphisms. The total field
content in the extended history phase space is thus
\bes
\barr{|l|lll|}    
\hline
\hbox{afn} & \hbox{Field} & \hbox{Momentum} & \hbox{Parity}\\
\hline
-1 & c^\mu_{diff}(x) & b_\mu^{diff}(x) & F\\
0 & g_{\mu\nu}(x) & \pi^{\mu\nu}(x) &B \\
1 & g^{\mu\nu}_*(x) & \pi^*_{\mu\nu} &F\\
2 & \zeta^\mu(x) & \chi_\mu(x) &B\\
\hline
\earr
\eens
The KT differential $\dlt$ is defined by
\bes
\dlt c^\mu_{diff}(x) &=& 0, \nl
\dlt g_{\mu\nu}(x) &=& 0, \nle
\dlt g^{\mu\nu}_*(x) &=&  G^{\mu\nu}(x) \nl
\dlt \zeta_\mu(x) &=& \nabla_\nu G^{\mu\nu}(x),
\eens
i.e. the KT operator is
\be
Q_{KT} = \int \dFx\ \ (G^{\mu\nu}(x)\pi^*_{\mu\nu}(x)
+ \nabla_\nu G^{\mu\nu}(x)\chi_\mu(x)).
\ee
The longitudinal operator was written down in (\ref{QLongC}),
$Q_{Long} = Q_{Long}^{diff}$, and the BRST operator is the sum of the
KT and the longitudinal operators, as usual.

We now quantize by passing to jet space, introducing a Fock
vacuum that is annihilated by the negative frequency modes of all fields
and antifields, and normal ordering. 
The fields are symmetric tensor fields, i.e. they correspond to the
symmetric $gl(N)$ modules $S_\ell$ and $S^\ell$ in (\ref{kSl}). 
The values of the parameters in (\ref{numdef}) are 
\bes
\barr{|l|l|c|cccc|c|}    
\hline
\afn & \hbox{Field} & \rep & u & v & w & x & p\\
\hline
-1 & c^\mu_{diff}(x) & S^1 & 1 & 0 & -1 & N & p-2 \\
0 & g_{\mu\nu}(x) & S_2 & -(N+2) & -1 & -(N+1) & -N(N+1)/2 & p  \\
1 & g^{\mu\nu}_*(x) & S^2 & (N+2) & 1 & -(N+1) & N(N+1)/2 & p-2 \\
2 & \zeta^\mu(x) & S^1 & -1 & 0 & 1 & -N & p-3 \\
\hline
\earr
\eens
The parameters were written down in arbitrary dimension $N$ for 
generality, although we are primarily interested in the physical case 
$N=4$. 
The last column is the truncation order for the corresponding jets.
That the antifields with $\afn = 1$ and $2$ should be truncated at order
$p-2$ and $p-3$, respectively, follow because Einstein's equation is
second order and the identity (\ref{Econt}) third order. That the ghost
contribution should be truncated at order $p-2$ is less clear, and
I have made a different assignment elsewhere \cite{Lar05a}. The reason why
I now favor $p-2$ is that this leads to nice cancellation of infinities.

The diffeomorphism anomalies are now read off from (\ref{cs}); for
definiteness, we only consider $c_1$. The abelian
charge $c_1 = 1 + c^{ghost}_1 + c^g_1 + c^{g*}_1 + c^\zeta_1$, where
\bes
c^{ghost}_1 &=& - {N+p-2\choose N} - N {N+p-1\choose N+2}, \nl
c^g_1 &=& (N+2){N+p\choose N} + {N(N+1)\/2}{N+p+1\choose N+2}, \nl
c^{g*}_1 &=& - (N+2) {N+p-2\choose N} 
- {N(N+1)\/2} {N+p-1\choose N+2}, \nl
c^\zeta_1 &=&  {N+p-3\choose N} + N {N+p-2\choose N+2}.
\label{c1grav}
\ee
It is clear that $c_1$ does not vanish for generic $p$.
The longitudinal operator (\ref{QLongJ})
thus acquires an anomaly, and we can only implement the KT cohomology.
Hence the ghost plays no direct role, and I have previously argued that
should be discarded. Nevertheless, it softens the divergence in the
$p\to\infty$ limit, since
\be
c^{ghost}_1 + c^\zeta_1  = - {N+p-3\choose N-1} - N {N+p-2\choose N+1}
\ee
only diverges like $p^{N+1}/(N+1)!$, while the leading term in 
(\ref{c1grav}) is proportional to $p^{N+2}/(N+2)!$.

The quantum KT operator becomes
\bes
Q_{KT} &=& \int dt\ \Big\{
\summ{p-2} \no{ G^{\mu\nu}_\cmm(t)\pi^{*\cmm}_{\mu\nu}(t) } \nle
&&+ \summ{p-3} \no{ (\nabla_\nu G^{\mu\nu})_\cmm(t)\chi^\cmm_\mu(t) } 
\Big\},
\eens
where $G^{\mu\nu}_\cmm(t)$ and $(\nabla_\nu G^{\mu\nu})_\cmm(t)$
are the corresponding jets.

\section{Finiteness conditions}
\label{sec:Finiteness}

In the previous section we applied the MCCQ formalism to gravity, and 
found a well-defined but anomalous action of the DGRO algebra. However,
the passage to the space of $p$-jets amounts to a regularization. The
regularization is unique in that it preserves the full constraint
algebra, but it must nevertheless be removed in the end. In order to
reconstruct the original field by means of the Taylor series
(\ref{Taylor}), we must take the limit $p\to\infty$. A necessary
condition for taking this limit is that the abelian charges have a
finite limit. 

Taken at face value, the prospects for succeeding
appear bleak. When $p$ is large, ${m+p\choose n} \approx p^n/n!$, so 
the abelian charges (\ref{cs}) diverge; the worst case is 
$c_1 \approx c_2 \approx p^{N+2}/(N+2)!$, which diverges in all dimensions
$N > -2$. In \cite{Lar01} a way out of this problem was devised: consider 
a more general realization by taking the direct sum of operators 
corresponding to different values of the jet order $p$. Take
the sum of $r+1$ terms like those in (\ref{Fock}), with $p$ replaced by
$p$, $p-1$, ..., $p-r$, respectively, and with $\rep$ and $M$ replaced by
$\repi$ and $\Mi$ in the $p-i$ term. 

Such a sum of contributions arises naturally from the KT and BRST
complexes, because
the antifields are only defined up to an order smaller than $p$ (e.g.
$p-1$ or $p-2$). 
Denote the numbers $u,v,w,x,y$ in the modules $\repi$ and $\Mi$,
defined as in (\ref{numdef}), by $u_i,v_i,w_i,x_i,y_i$, 
respectively. Of course, there is 
only one contribution from the observer's trajectory. 
Then it was shown in \cite{Lar01}, Theorem 3, that
\bes
c_1 = -U\Npr, &\qquad&
c_2 = -V\Npr, \nl
c_3 = W\Npr, &\qquad&
c_4 = -X\Npr,
\label{finc} \\
c_5 = Y\Npr, 
\eens
where $u_0=U$, $v_0=V$, $w_0=W$, $x_0=X$ and $y_0=Y$, provided that the 
following conditions hold:
\bes
&&u_i + \ritwo X = \mri U, \nl
&&v_i - 2\rione W - \ritwo X = \mri V, \nl
&&w_i - \rione X = \mri W, 
\label{conds}\\
&&x_i = \mri X, \nl
&&y_i = \mri Y.
\eens
The contributions from the observer's trajectory have also been eliminated
by antifields coming from the geodesic equation; this is
not important in the sequel because these contributions are finite 
anyway.

Let us now consider the solutions to (\ref{conds}) for the numbers
$x_i$, which can be interpreted as the number of fields and anti-fields.
First assume
that the field $\fam(t)$ is fermionic with $x_F$ components, which gives
$x_0=x_F$. We may assume, by the spin-statistics theorem, that the 
EL equations are first order, so the bosonic antifields $\fsam(t)$ 
contribute $-x_F$ to $x_1$. The barred antifields $\wfam(t)$ are also
defined up to order $p-1$, and so give $x_1=-x_F$, and the barred 
second-order antifields $\wfsam(t)$ give $x_2=x_F$. 

For bosons the situation is analogous, with some exceptions: all signs
are reversed, and the EL equations are assumed to be second order. Hence
$\fam(t)$ yields $x_0=-x_B$ and $\fsam(t)$ yields $x_2=x_B$. 
Moreover, we have the second-order
antifields $\zam(t)$ which contribute $x_3=-x_G$, and the ghosts
$\cam(t)$. We assume, only because this choice will lead to
cancellation of unwanted terms, that the ghosts are truncated at order
$p-2$, and thus they contribute $x_2=x_G$. 
Again, there are barred antifields of opposite parity at one order higher.

Previously I have argued \cite{Lar02} that the fermionic EL equations
should also have $x_S$ gauge symmetries, i.e. there are also fermionic
second-order antifields $\zam(t)$ which give $x_2=x_S$. The motivation
for this assumption was that a non-zero value for $x_S$ is necessary to
cancel some infinitites in the $p\to\infty$ limit. However, $x_S=0$ in
all established theories (a non-zero $x_S$ may be thought of as some kind
of supersymmetry), so this assumption was a major embarrassment. Because
it turns out that infinities cancel precisely when $x_S = x_G$, the role
of these hypothetical fermionic second-order antifields can be played by
the ghosts. This is the reason why I now favor that the ghosts be kept,
despite $Q^2_{BRST}\neq 0$, and that they should be truncated at one
order higher than $\zam(t)$.

The situation is summarized in the following tables, where the upper half
is valid if the original field is fermionic and the lower half if it is
bosonic:
\bes
\barr{|c|c|c|l|}                                                       
\hline
\afn & \hbox{Jet} & \hbox{Order} &x \\
\hline
0 & \fam(t) & p & x_F \\
1 & \wfam(t) & p-1 & -x_F \\
1 & \fsam(t) & p-1 & -x_F \\
2 & \wfsam(t) & p-2 & x_F \\
\hline
\hline
0 & \fam(t) & p & -x_B \\
1 & \wfam(t) & p-1 & x_B \\
1 & \fsam(t) & p-2 & x_B \\
2 & \wfsam(t) & p-3 & -x_B \\
-1 & \cam(t) & p-2 & x_G \\
0 & \bar\cam(t) & p-3 & -x_G \\
2 & \zam(t) & p-3 & -x_G \\
3 & \bar\zam(t) & p-4 & x_G \\
\hline
\earr
\label{tab}
\ees
If we add all contributions of the same order, we see that the fourth 
relation in (\ref{conds}) can only be satisfied provided that
\bes
p: &\quad& x_F - x_B = X \nl
p-1: && -2x_F+x_B = -rX, \nl
p-2: && x_B + x_F + x_G = {r\choose2}X, \nl
p-3: && -x_B-2x_G = -{r\choose3}X, 
\label{rcond}\\
p-4: && x_G = {r\choose4}X, \nl
p-5: && 0 = -{r\choose5}X, ...
\eens
The last equation holds only if $r\leq4$ (or trivially if $X=0$). 
On the other hand, if we demand that there is at least one bosonic
gauge condition, the $p-4$ equation yields $r\geq4$. Such a demand is
natural, because both the Maxwell/Yang-Mills and the Einstein equations
have this property. Therefore, we are unambigiously guided to consider
$r=4$ (and thus $N=4$). The specialization of (\ref{rcond}) to four
dimensions reads
\bes
p: &\quad& x_F - x_B = X \nl
p-1: && -2x_F+x_B = -4X, \nl
p-2: && x_B + x_F + x_G = 6X, \\
p-3: && -x_B-2x_G = -4X, \nl
p-4: && x_G = X.
\eens
Clearly, the unique solution to these equations is
\be
x_F = 3X, \qquad x_B = 2X, \qquad x_G = X.
\label{xsol}
\ee
The solutions to the remaining equations in (\ref{conds}) are found by
analogous reasoning. The result is
\bes
\barr{llllll}
u_B = 2U  &\qquad& v_B = 2V+2W \\
u_F = 3U  && v_F = 3V+2W \\
u_G = U-X && v_G = V+2W+X \\
\label{uvwysol} \\
w_B = 2W+X && y_B = 2Y \\
w_F = 3W+X && y_F = 3Y \\
w_G = W+X  && y_G = Y \\
\\
\earr
\ees
This result expresses the fifteen parameters $x_B-w_G$
in terms of the five parameters $X$, $Y$, $U$, $V$, $W$.
For this particular choice of parameters, the abelian charges in
(\ref{finc}) are given by 
\be
c_1 = -U, \qquad c_2 = -V, \qquad c_3 = W, \qquad c_4 = -X,
\qquad c_5 = Y,
\ee
independent of $p$. Hence there is no manifest obstruction to the
limit $p\to\infty$.

\section{ Comparison with known physics }
\label{sec:Known}

The analysis in the previous section yields highly non-trivial constraints
on the possible field content, if we demand that it should be possible
to remove the jet regularization. It is therefore interesting to compare
the the numbers in (\ref{xsol}), with the field content of gravity 
coupled to the standard model in four dimensions. This is a slightly
modified form of an analysis given in \cite{Lar02,Lar03}.

The bosonic content of the theory is given by the following table. Standard
notation for the fields is used, and one must remember that it is the 
na\"\i ve number
of components that enters the equation, not the gauge-invariant physical
content. E.g., the photon is described by the four components $A_\mu$ 
rather than the two physical transverse components.
Also, the gauge algebra $\ssu$ has $8+3+1 = 12$ generators.
\bes
\barr{|c|c|c|c|}
\hline
\hbox{Field} & \hbox{Name}& \hbox{EL equation} & x_B  \\
\hline
A^a_\mu & \hbox{Gauge bosons}
&D_\nu F^{a\mu\nu} = j^{a\mu} &12\times4 = 48  \\
g_{\mu\nu} & \hbox{Metric}
&G^{\mu\nu} = {1\/8\pi}T^{\mu\nu} & 10 \\
H & \hbox{Higgs field}
& g^{\mu\nu}\dmu\dnu H = V(H) & 2\\
\hline
\earr
\nle
\barr{|c|c|}
\hline
\hbox{Gauge condition} & x_G  \\
\hline
D_\mu D_\nu F^{a\mu\nu} = 0 & 12\times1 = 12 \\
\nabla_\nu G^{\mu\nu} = 0 & 4 \\
\hline
\earr
\eens
The total number of bosons in the theory is thus $x_B = 48+10+2 = 60$,
which implies $X=30$ by (\ref{xsol}). The number of gauge conditions is
$x_G = 16$, which implies $X=16$. There is certainly a discrepancy here.

The fermionic content in the first generation is given by
\bes
\barr{|c|c|c|c|}
\hline
\hbox{Field} & \hbox{Name}& \hbox{EL equation} & x_F  \\
\hline
u & \hbox{Up quark} & \Dslash u = ... & 2\times3 = 6\\
d & \hbox{Down quark} & \Dslash d = ... & 2\times3 = 6\\
e & \hbox{Electron} & \Dslash e = ... & 2\\
\nu_L & \hbox{Left-handed neutrino} & \Dslash \nu_L = ... & 1\\
\hline
\earr
\ees
The number of fermions in the first generation is thus 
$x_F = 6+6+2+1 = 15$. Counting all three generations and anti-particles,
we find that the total number of fermions is $x_F = 2\times3\times15 = 90$,
which implies $X=30$. 

The prediction that spacetime has $N=4$ dimensions is of course nice,
but the values for $X$ ($30,16,30$) are not mutually consistent. 
One could think of various modifications. By identifying $x_i$ with 
the number of field components, we have assumed that the weight $\la=0$
in (\ref{jet}). A non-zero $\la$ would modify the abelian charge.
Unfortunately, this is only possible for the fermions which obey linear 
equations of motion, so the discrepancy $x_B \neq 2x_G$ remains.
Another problem, found in \cite{Lar05a}, is that the gauge anomaly 
$c_5$ does not have a finite $p\to\infty$ limit for reasonable choices 
of field content.

Hence it is presently unclear how to remove the regulator and take the
field limit, and this is of course a major unsolved problem. Nevertheless,
it should be emphasized that already the regularized theories carry
representations of the {\em full} gauge and diffeomorphism algebras.
One may also hope that some overlooked idea may improve the situation,
as happened with the inclusion of ghosts.

\section{Conceptual issues}
\label{sec:Concept}

One of the most important tasks of any putative quantum theory of
gravity is to shed light on the various conceptual difficulties which
arise when the principles of quantum mechanics are combined with
general covariance \cite{Car01,Sav04}. These issues include:
\begin{enumerate}
\item
In conventional canonical quantization, the canonical commutation
relations are defined on a ``spacelike'' surface. However, a surface is
spacelike w.r.t. some particular spacetime metric $g_{\mu\nu}$, which is
itself a quantum operator.
\item
Microcausality requires that the field variables 
defined in spacelike separated regions commute. Again, it is unclear
what this means when the notion of spacelikeness is dynamical.
\item
Different choices of foliation lead to a priori different quantum 
theories, and it by no means clear that these are unitarily equivalent.
\item
The problem of time: The Hamiltonian of general relativity is a first
class constraint, hence it vanishes on the reduced phase space. This 
means that there is no notion of time evolution among 
diffeomorphism-invariant degrees of freedom.
\item
The arrow of time: Dynamics is invariant under time reversal (combined
with CP reversal), but we nevertheless experience a difference between
past and future.
\item
The notion of time as a causal order is lost. This is not really a
problem in the classical theory, where one can solve the equations of
motion first, but in quantum theory causality is needed from the outset.
\item
QFT rests on two pillars: quantum mechanics and locality. However, 
locality is at odds with diffeomorphism invariance underlying gravity;
``there are no local observables in quantum gravity''.
\end{enumerate}

Let us see how MCCQ addresses these conceptual issues.
\begin{enumerate}
\item
The canonical commutation relations are defined throughout the history
phase space $\PP$, and hence not restricted to variables living on a
spacelike surface. Dynamics is implemented as a first class constraint
in $\PP$. Only if we solve this constraint prior to quantization need we
restrict quantization to a spacelike surface.
\item
By passing to $p$-jet space, we eliminate the notion of spacelikeness
altogher. The $p$-jets live on the observer's trajectory, and the 
observer moves along a timelike curve. It
might seem strange to dismiss the notion of spacelike separation, but
distant events can never be directly
observed, and a physical theory only needs to describe directly 
observable events. What can be observed are indirect effects of distant
events. E.g., a terrestial detector does not directly observe the sun, 
but only photons emanating from the sun eight light-minutes ago. The 
detector signals are of course compatible with the existence of the sun, 
but a physical theory only needs to deal with directly observed events,
i.e. the absorbtion of photons in the detector.
\item
In MCCQ there is no foliation, but rather an explicit observer, or 
detector. The theory is
unique since the observer's trajectory is a quantum object; we do not
deal with a family of theories parametrized by the choice of observer,
but instead the observer's trajectory is represented on the Hilbert 
space in the same way as the quantum fields.
\item
By introducing an explicit observer, we can define a genuine energy
operator (\ref{Hq}) which translates the fields relative to the observer,
or vice versa. In contrast, there is also a Hamiltonian constraint, which
translates both the observer and the fields the same amount. This
constraint is killed in KT cohomology and is thus identically zero on
physical observables.
\item
The Fock vacuum $\ket0$ (\ref{LER}) treats positive and negative energy 
modes differently, thus introducing an asymmetry between forward and 
backward time translations.
\item
The $p$-jets live on the observer's trajectory $q(t)$ and are
thus causally related; causal order is defined by the parameter $t$. 
The relation between this order and the fields is encoded in the
geodesic equation.
\item
As we saw in Section \ref{sec:Locality}, locality is compatible with
infinite-dimensional spacetime symmetries, but only in the presence of 
an anomaly. This is the key lesson from CFT.
\end{enumerate}
It is gratifying that the MCCQ formalism yields natural explanations of
many of the conceptual problems that plague quantum gravity.

\section{Conclusion}
\label{sec:Conclusion}

The key insight underlying the present work is that the process of
observation must be localized in spacetime in order to be compatible with
the philosophy of QFT. The innocent-looking introduction of the
observer's trajectory leads to dramatic consequences, because new gauge
and diffeomorphism anomalies arise. On the mathematical side, this
construction leads to well-defined realizations of the constraint algebra
generators as operators on a linear space, as least for 
the regularized theory.

We have also further developed the manifestly covariant canonical
quantization method, based on the form of the DGRO algebra modules, which
was introduced in \cite{Lar04,Lar05a}. This formalism is convenient due
to its relation to representation theory, but it is presumably possible
to repeat the analysis in any sensible quantization scheme, at the cost
of additional work. In contrast, the introduction of the observer's
trajectory is absolutely crucial, because the new anomalies can not be
formulated without it. Anomalies matter!

Four critical problems remain to be solved. As was discussed in Section
\ref{sec:Finiteness}, the original fields must be reconstructed from the
$p$-jets, i.e. we must take the limit $p\to\infty$. This limit is
problematic because the abelian charges diverge. Second, the issue of
unitarity needs to be understood. So far we only noted that an extension
is necessary for unitarity by restriction to Virasoro subalgebras, and
then we proceeded to construct anomalous representations. The main
problem is to find an invariant inner product. Third, perturbation
theory and renormalization must be transcribed to ths formalism, to make
contact with numerical predictions of ordinary QFT. Finally, we know
from CFT that reducibility conditions analogous to Kac' formula
\cite{FMS96} are needed in physically interesting situations.
Unfortunately, none of these problems appears to be easy.

\end{document}